\documentclass{jfm}
\usepackage{lmodern}
\usepackage{pdftexcmds}
\usepackage{amsmath}
\usepackage{bm}

\usepackage{graphicx}
\usepackage{epstopdf, epsfig}
\usepackage{natbib}

\usepackage{epsfig}
\usepackage{amsmath}
\usepackage{amsfonts}
\usepackage{amssymb}
\usepackage{wrapfig}
\usepackage{pgf}
\usepackage{caption}
\usepackage{subcaption}
\usepackage{cite}

\usepackage{epstopdf}
\usepackage{algorithm}
\usepackage{algpseudocode}

\usepackage{xcolor}
\definecolor{khaki}{rgb}{0.8,0.7,0.3}
\definecolor{Mycyan}{rgb}{0.,1.,1.}
\definecolor{Mypink}{rgb}{1.,0.,1.}
\definecolor{Myyellow}{rgb}{1.,1.,0.}
\usepackage{tikz}
\usetikzlibrary{shapes}

\usepackage{graphbox}

\usepackage{parskip}
\usepackage{enumitem}

\DeclareMathAlphabet{\mathsfbi}{OT1}{\sfdefault}{bx}{sl}
\DeclareMathVersion{sfletters}
\SetSymbolFont{letters}{sfletters}{OML}{ntxsfmi}{b}{it}
\makeatletter
\graphicspath{{Figures/}}

\newcommand{\mathbfsbilow}[1]{%
  \text{\mathversion{sfletters}$\m@th#1$}%
}
\DeclareRobustCommand{\tensor}[1]{%
  \begingroup
  \ifcat\noexpand #1\relax
    \mathbfsbilow{#1}%
  \else
    \mathsfbi{#1}%
  \fi
  \endgroup
}
\makeatother

\shorttitle{Viscoplastic flow with reaction}
\shortauthor{M. Zare,  I.A.~Frigaard and G.~Lawrence}

\title{Bubble-Induced Entrainment at Viscoplastic-Newtonian Interfaces}
\author[M. Zare, I.A. Frigaard \& G. Lawrence]
{M.\ns Z\ls A\ls R\ls E$^{1, \thanks{Email address for correspondence: marjan.zare@mech.ubc.ca}}$,
I.\ns A.\ns F\ls R\ls I\ls G\ls A\ls A\ls R\ls D$^{1,2}$
  \and  G.\ns L\ls A\ls W\ls R\ls E\ls N\ls C\ls E$^3$}

\affiliation{%$^1$Department of Applied Mathematics and Theoretical Physics, University of Cambridge, Wilberforce Road, Cambridge, CB3 0WA, UK.

$^1$Department of Mathematics, University of British Columbia, Vancouver, 
Canada.
$^2$Department of Mechanical Engineering,
 University of British Columbia, Vancouver, Canada.
$^3$Department of Civil Engineering,
 University of British Columbia, Vancouver, Canada.
}
\pubyear{2020}
\volume{?}
\pagerange{?}

\date{?; revised ?; accepted ?. - To be entered by editorial office}

\begin{document}

\maketitle

\begin{abstract}
The passage of single air bubbles through the horizontal interface between miscible viscoplastic and Newtonian fluids, considering various combinations of densities and viscosities for the fluid layers, is studied computationally. The primary focus is on the quantity of liquid transferred from the lower layer (Viscoplastic fluid) to the upper layer (Newtonian fluid) as a result of the bubble's ascent, a factor with significant implications for the turbidity of methane-emitting lakes and water bodies. The results show that at $ Bo>1 $ and moderate $ Ar $, prolate-shaped bubbles crossing the interface undergo elongation in the direction of their poles. This elongation is further accentuated when the viscosity of upper layer is less than the plastic viscosity of the lower layer. The bubble is found to break up when leaving the lower layer, of a critical capillary number, $ Ca_c \approx 5 $. The results show a significant reduction in the volume of entrainment compared to the Newtonian counterpart. This suggests disturbances caused by the rising bubble at the interface dissipate over a smaller region. Four distinct entrainment regimes are identified, mainly indicating the height to which the entrained fluid can be transported away from the interface. In contrast to Newtonian fluids, the volume of entrainment increases by decreasing the viscosity of the upper layer. Interestingly, the heavy yield stress fluid that has been dragged up into the the light Newtonian fluid does not recede down by time. 
\end{abstract}

\begin{keywords}
Viscoplastic fluids, entrainment, rising bubbles, bubbles break up. 
\end{keywords}

\section{Introduction}
\label{sec:intro}

Turbidity, defined as the cloudiness or opaqueness of water, serves as a critical indicator of the overall health and dynamics of aquatic ecosystems such as tailing ponds. It provides insights into the presence and distribution of suspended particles, the penetration of sunlight, and the consequent absorption of solar energy. This, in turn, directly influences the ecological equilibrium and the biological productivity of the aquatic ecosystem. Additionally, turbidity significantly contributes to the visual aesthetics of aquatic environments and serves as a reliable indicator of water quality, often revealing the presence of pollutants or sediments. As we grapple with industrial challenges such as managing tailings, and environmental challenges such as climate change and land and water resource management, a thorough understanding of turbidity is essential for informed decision-making, conservation efforts, and the sustainable preservation of our precious land and water resources.

Base Mine Lake (BML) is a man-made water body in Alberta, Canada. The lake is the first full scale demonstration end pit lake in the region and consists of water with an average depth of 9 m, covering a layer of fluid fine tailings (FFT) of up to 45 m deep. FFT includes elements such as dissolved organic matter, suspended organic, and inorganic particles that can collectively contribute to turbidity. Examination of suspended solids concentrations in the water cap above the FFT suggested oscillating bottom currents beneath wind-waves as a potential contributor to FFT erosion \citep{lawrence1991wind}. However, subsequent analysis by the authors has led to the conclusion that this is unlikely to serve as a significant direct source of turbidity in Base Mine Lake. The water cap's substantial depth in Base Mine Lake renders these currents, which attenuate exponentially with increasing depth, too weak to erode the FFT. Echo soundings conducted beneath the ice at BML have revealed the ascent of gas bubbles through the water column, as depicted in Fig. 4 of \citet{lawrence2016suspended}. These observations, combined with water turbidity data measured during winter, as reported in \citet{tedford2019temporal}, give rise to the hypothesis that these bubbles are likely facilitating the transportation of some solids from the FFT into the water cap, while also potentially inducing circulation and mixing within the water cap. The primary objective of the present study is to examine the entrainment of fluid from the lower layer to the upper layer, by rising bubbles.

The lower layer in which bubbles are produced (FFT layer) has been characterised as a yield stress fluid \citep{Derakhshandeh2016}. In such fluids the material flows only if the imposed stress surpasses the yield stress \citep{Balmforth2014}. In the context of bubbles rising in a viscoplastic fluid, one might expect that stresses arise from both surface tension and buoyancy effects, and are resisted by the yield stress of the material. Finding the shape and position of the yield surfaces, i.e. the boundaries between the yielded and unyielded regions, is non-trivial. This complicates understanding bubble migration through yield stress fluids, and further the entrainment process when one layer features yield stress. 

During the last decades, many theoretical and computational studies have been performed focused at determining the yield surfaces around moving objects in yield stress fluids \citep{tsamopoulos2008steady,dimakopoulos2013steady,sikorski2009motion, lopez2018rising,chaparian2021clouds,Emad2022,daneshi2023growth,kordalis2023hydrodynamic}.  Theoretical and computational studies amongst these are often based on model of ideal (or simple) yield stress fluids, as defined in \citet{frigaard2019simple}. Experimentally, deviation from ideal viscoplastic behaviour is always present. Experimental observations find fore-and-aft asymmetries in moving bubbles, even at low Reynolds numbers, characterized by an inverted teardrop shape of the bubble as it rises and a negative wake at the rear of the bubble \citep{sikorski2009motion,mougin2012, lopez2018rising,Zare2018INV2,pourzahedi2021,daneshi2023growth}. These features are attributed to viscoelastic behaviour of the fluid around the bubble \citep{moschopoulos2021concept}. These same viscoelastic stresses result in successive released bubbles following the same pathway \citep{lopez2018rising}. These \emph{damaged} or \emph{weakend} pathways appear relevant to bubble migration in large reservoirs. \citet{ZareJFM1} showed that the shape and trajectory of bubbles are significantly influenced when they move within a distance (not greater than their yielded zone) from a \textit{weaker} region. This influence arises from alterations in residual stress distributions. 

In the context of entrainment, we are, in particular, interested in the flow around the bubble near viscoplastic-Newtonian interface. We conducted a series of experiments to investigate the behavior of bubbles rising in a two-layer system consisting of viscoplastic water, as detailed in \citet{zhao2022bubbles}. Our findings revealed that as the bubble enters the water layer, it entrains viscoplastic fluids, which accumulate above the interface in a conical shape. Additionally, we observed the formation of water conduits extending downward into the lower layer, when multiple bubbles passed through. As the bubbles ascended through these conduits, they displaced water, and once they exited, water returned to fill the conduits. This phenomenon holds significant environmental significance, as it can facilitate the movement of contaminants to and from submerged sediments. To the best of our knowledge, there is no other research that explores the ascent of bubbles through a rheologically non-uniform yield stress medium. 
 
The concept that a moving particle permanently displaces the surrounding fluid was quantified by \citet{darwin1953note}. Early investigations have been based on this classical concept, known as Darwin's drift. In these studies, a pseudo-three-phase system is modelled, in which the two fluid layers consist of a single liquid, and the entrainment of fluid by body translation is predicted in Stokes regime \citep{eames1997displacement,bush1998fluid}. Numerous experimental and numerical investigations have been conducted to explore the process of bubbles passing through a Newtonian liquid-liquid interface since then \citep{mercier1974influence,greene1988onset,greene1991bubble,
reiter1992observations,manga1995low, kemiha2007passage,dietrich2008passage,uemura2010ripples,bonhomme2012inertial,natsui2014multiphase,emery2018flow}. \citet{greene1988onset,greene1991bubble} established a crude criterion for bubble penetration of the liquid-liquid interface by comparing buoyancy and surface tension between the liquid layers. Their work identified the conditions under which a bubble can be trapped at the interface. They noted that the entrainment volume can be eliminated by increasing the density ratio between the two liquids. Interestingly, they observed that interfacial tension had a relatively modest impact on the entrainment volume but played a role in initiating entrainment. Furthermore, they found that a decrease in the viscosity of the lower liquid resulted in a significant increase in the entrainment volume. 

\citet{reiter1992observations,reiter1992characteristics} conducted a study in which they recorded the duration of the bubble's presence at the interface (residence time), the vertical extent of the column formed beneath the bubble, and the properties of drops formed in the upper phase. \citet{shopov1992unsteady} conducted transient numerical simulations at low and intermediate Reynolds number, to study bubbles crossing the interface between immiscible fluids. Specifically focusing on bubble and interface deformations as well as film drainage dynamics, without considering the tail-pinch off phase. Their findings revealed that at very low Weber and Reynolds numbers, bubbles could take on a prolate shape (elongated in the direction of motion) when passing through the interface, particularly in cases that the upper layer is less viscous, a result supported by \citet{manga1995low}. In contrast, at higher Weber and Reynolds numbers, inertial forces and interaction with the interface cause bubbles to adopt oblate shapes, often forming a concavity at the rear and spherical cap shapes during the crossing.

\citet{dietrich2008passage} experimentally investigated interface crossing in inertial regimes, establishing a relationship between the crossing time of the interface by a bubble and the ratio of terminal velocities in the two continuous phases. They observed the coexistence of fluid motion entrained in the bubble wake and an opposing flow driven by gravity once the tail formed. \citet{bonhomme2012inertial} considered bubble crossing the interface between immiscible fluids at inertial conditions, examining a wide range of bubble shape configurations from spherical to toroidal. They provided a comprehensive map of bubble shapes and entrained column geometries based on Archimedes and Bond numbers. Their findings indicated that smaller bubbles could be slowed down or even stopped at the interface, whereas larger spherical cap bubbles, with larger cross-sectional areas, generally crossed the interface more easily. \citet{emery2018flow} explored the crossing of single bubbles considering the tail-pinch off mode and found that the length of the entrained liquid column was longer when the bubble's velocity remained relatively constant during the interface crossing. The tail often remained connected for an extended period before rupturing, with the liquid shell covering the bubble breaking before the column. They also studied the crossing of a stream of bubbles, providing insights into various flow regimes and the potential formation of clusters.

The majority of previous work on bubble passage through liquid-liquid interface reviewed has focused solely on Newtonian fluids. However, the passage of a single bubble through a viscoplastic-Newtonian fluid has not been studied before to our knowledge and very little is known about bubble's motion in a rheologically non-unifom fluids. This paper embarks on a comprehensive exploration of the relationship between the elementary physical mechanisms at play in the entrainment of yield stress fluid by the rising of a single bubble, using computational method. The study does not focus specifically on any one aspect but instead intends to provide a broad characterization and overview of the possible flow regimes. The following sections describe the computational methods, provide technical details, and discuss the results in depth, offering valuable insights into the dynamics of bubbles at fluid-fluid interfaces and subsequent entrainment.

\section{Problem Formulation}

\begin{figure}
\centering
\includegraphics[scale=1.1]{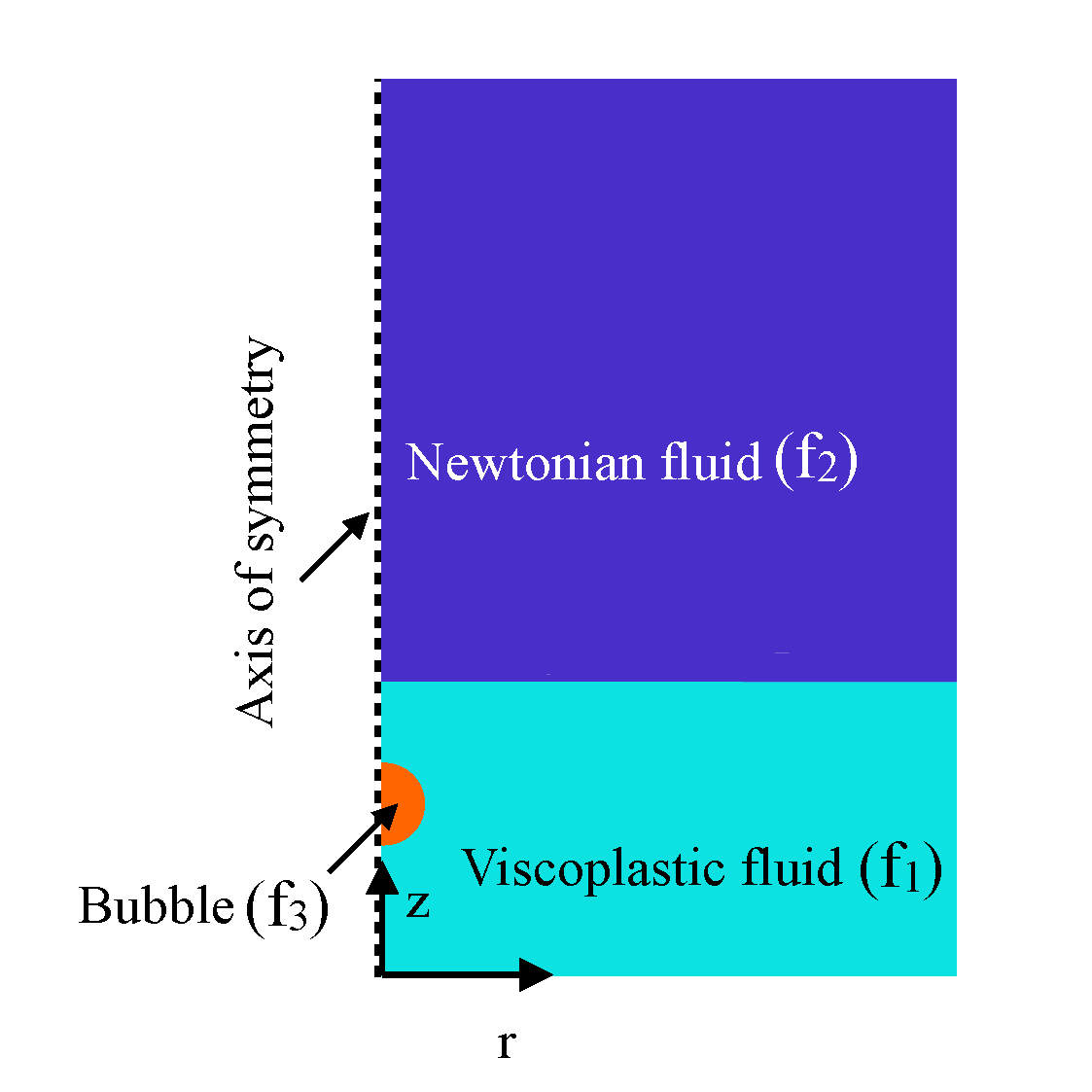}
%\vspace{-2.5cm}
\caption{Schematic of the flow geometry and coordinate system.}
\label{fig:Geom}
\end{figure}

The present study is centered on the flow around a single gas bubble crossing an interface between two liquid layers. Our primary focus revolves around entrainment, which refers to an amount of liquid from the lower layer that is transported to the upper layer by the rising bubble. Fig.~\ref{fig:Geom} illustrates the flow geometry examined herein.

 Our simulations are computed in an axisymmetric domain ($ [0,100 \hat{R}_b] \times [-15\hat{R}_b, 50\hat{R}_b])$ in $ (\hat r,\hat z) $ coordinates. Here $ \hat{R}_b $ denotes the radius of the initially spherical bubble. The liquids are initially static separated with a horizontal interface at $\hat z=0 $. The bubble is initially placed in the lower part of the flow domain at $ \hat z=-5 $.

 The flow surrounding a rising bubble is governed by the usual momentum and mass balances, and we make them dimensionless through the following approach. We scale all lengths with the equivalent radius, $ \hat{R}_b $, of a spherical bubble of the same volume. We scale velocities with $ \hat U$, obtained by balancing buoyancy and viscous forces, i.e.~$ \hat U=\hat{\rho}_1\hat{g}\hat{R}_b^2/\hat{\mu}_p $, where $ \hat{g} $ is the gravitational acceleration and $ \hat{\mu}_p $ is the plastic viscosity of the surrounding viscoplastic fluid. Pressure and other stresses are scaled with $\hat{\rho}_1\hat{g}\hat{R}_b$.
 
The dimensionless groups that emerge are: Archimedes number ($Ar=\hat{\rho}_1^2\hat{R}_b^3\hat{g}/\hat{\mu}_{p}^2$), representing gravitational forces to the effects of effective viscosity, the Bond number ($ Bo=\hat{\rho}_1\hat{g}\hat{R}_b^2/\hat{\sigma}_s$), representing the ratio gravitational forces to surface tension effects, where $ \hat{\sigma}_s$ is the surface tension coefficient, the density ratio ($ \rho $), representing the relationship between the density of the upper layer and the density of the lower layer, note that air density is negligible, $\rho = \rho_b  \ll 1$, yield number ($Y=\hat{\tau}_{Y}\hat{R}_b/\hat{\rho}_1\hat{g}$) representing the competition between yield and buoyancy stresses, and viscosity ratio ($ m=\hat{\mu}_2  / \hat{\mu}_1$), note that $m_b = \hat{\mu}_b/\hat{\mu}_p \ll 1$.

The scaled momentum and continuity equations are given by
\begin{equation} \label{eq:Momentum}
\rho Ar [\boldsymbol{u}_t+\boldsymbol{u} \cdot \nabla\boldsymbol{u}]=-\nabla p + \mathbf{\nabla}  \cdot \boldsymbol \tau + \frac{k \boldsymbol n \delta_s}{Bo}  + \rho \boldsymbol{e_{g}}
\end{equation}
\begin{equation} \label{eq:Continuity1}
	\mathbf{\nabla} \cdot \boldsymbol u=0
\end{equation}
where $\boldsymbol u,~p,~\boldsymbol\tau$ denote the velocity, pressure and deviatoric stress, respectively. 

A volumetric force formulation has been adopted for the surface tension, and this term appears in the right hand side of the Navier-Stokes equation, eq. \ref{eq:Momentum}, where $ k $ is the mean curvature, $ \boldsymbol n $ the normal unit vector and $\delta_s$ is a surface Dirac $\delta$-function that is nonzero only on the interface. { Dimensionally this term is $   \hat{\mathbf{f}}= \hat{\sigma}_s k \bm n \delta_s $, where $\bm n \delta_s $ is approximated from $ \bm \nabla C$, which is the gradient of volume fraction. In the numerical scheme, this term is evaluated using height functions \citep{Popinet2009}.}

The upper layer fluid is treated as a Newtonian fluid, while the lower layer fluid is considered to be either a Newtonian or a yield stress fluid. Where we have Newtonian liquid in our flows, this is modelled by
\begin{equation} \label{eq:const-dpling}
	\boldsymbol{\tau}(\boldsymbol u)=m {\boldsymbol{\dot{\gamma}}}(\boldsymbol{u}),
\end{equation}

 where $\dot{\gamma}(\boldsymbol{u})$ is the strain rate tensor:
\begin{equation}
	{\boldsymbol{\dot{\gamma}}}(\boldsymbol{u})= \nabla \boldsymbol u +(\nabla \boldsymbol{u})^{T} .
\end{equation}

For the viscoplastic lower fluid layer the Bingham model has been used, which has been solved computationally using a commonly used smooth regularization method proposed by Papanastasiou \citep{papanastasiou1987}:

\begin{eqnarray}
\boldsymbol{\tau}(\boldsymbol u)=\left[1+Y\frac{1-e^{-N |\dot{\gamma}(\boldsymbol{u})|} }{|{\dot{\gamma}}(\boldsymbol{u})|}\right]{\boldsymbol{\dot{\gamma}}}(\boldsymbol{u}),
\label{eq:papanastasiou}
\end{eqnarray}

Here $\hat{\tau}_{Y}$ denotes the yield stress of the Bingham fluid. The regularization parameter $ N \gg 1$ controls the closeness of approximation to the exact Bingham fluid model: $1/N$ represents a small (scaled) strain rate below which the fluid becomes very viscous, i.e.~for $N |\dot{\gamma}(\boldsymbol{u})| \ll 1$, we find at leading order (\ref{eq:papanastasiou}) becomes
\[ \boldsymbol{\tau}(\boldsymbol u) \sim NY {\boldsymbol{\dot{\gamma}}}(\boldsymbol{u}) . \]
In dimensional terms the strain rate and $1/N$ have been scaled with $\hat U /\hat{R}_b = \hat{\rho}_1\hat{g}\hat{R}_b / \hat{\mu}_1 $.

Both fluids are considered miscible with density ratios of ($\rho = 1$) and ($\rho = 0.7$). Thus, there is no additional interfacial term required in (\ref{eq:Momentum}). Although miscible, the experimental timescale and length-scales are such that the P\'{e}clet number between liquids is extremely large and molecular diffusive effects can be neglected, i.e.~the VoF model is physically appropriate.

 The bubble density is fixed at $\rho_b = 0.001$. $ Ar $ is defined using the viscosity of the lower layer fluid. Therefore, the bubble viscosity ratio $m_b$ varies with the selected $Ar$. Nevertheless, for all $m_b$ studied we have $m_b \ll 0.002$. 
 
 Boundary conditions for the simulations are no-slip on the walls at the side and top of the domain. The pressure as well as the $z$-derivative of the normal velocity is set to zero tt the bottom of the domain ($z=-15$). Within the range of our dimensionless parameters, the size of yield surfaces develoing around the bubbles is expected not to exceed five times greater than the size of the bubble; see \citep{ZareJFM1}. The width of the domain is significantly larger than the width of yielded zone and the height of the domain is tall enough to allow for transients of interest to be captured.

 In steady state, the drag force on the bubble is balanced by the buoyancy. The drag coefficient $ C_d $ compares the drag force to the inertial force on the bubble:

\begin{equation} \label{eq:Cd}
C_d=\frac{32\hat{R}^3_b\hat{g}}{3\hat{U}^2_b\hat{W}^2}
\end{equation}

where $ \hat{W} $ denotes, the width of the steadily rising bubble.

\section{Computational method}

The computation has been performed using a Volume of Fluid (VoF) method. Equations (\ref{eq:Momentum}) \& (\ref{eq:Continuity1}) remain valid throughout the flow. The quadtree adaptive mesh refinement technique has been used to accurately track the interfaces between the two- and three-phases. The local mesh refinement densifies the grid in regions of strong spatial variations of the velocity, fluid concentrations and the strain rate $|\dot{\gamma}(\boldsymbol{u})|$. The algorithm is implemented in the open source multi-phase flow solver Basilisk (based on the algorithms of the former Gerris), which is specifically suited to such applications \citep{Popinet2003,Popinet2009,Popinet2019}. The system of equations is resolved using a time-splitting projection method described in \citep{Popinet2009}. A physically motivated maximum timestep is imposed and further controlled via a CFL constraint. The velocity advection term is estimated using the Bell-Colella-Glaz second-order upwind scheme at the intermediate timestep. The values of viscosity, density and pressure at the intermediate timestep are also used.

Basilisk employs the interpolation error between a field value at a grid point on grid level n and its interpolated value from a coarser grid level n-1 as a criterion to decide whether to adjust the grid locally. This adjustment involves either merging four squares into a parent square or dividing a square into four sub-squares in 2D, as appropriate. The grid is refined or coarsened depending on whether this interpolation error is larger or lower than $ 10^{-3}, 10^{-3}, 10^{-9} $ for  velocity, concentrations and strain rate, respectively. The domain is initially divided into $128 \times 128$ cells. Each of these may subdivided according to the refinement, to a maximum of 9 quadtree levels. Spurious currents are also minimized by using a height function method to estimate the interface curvature. Basilisk has been widely validated against analytical, numerical and experimental interfacial flows and has been frequently used for Newtonian bubble flow calculations \citep{Balla2019,Zhang2020,Berny2020}.

\subsection{Validation}

\begin{figure}
\hspace*{-0.1\textwidth}
\centering
\includegraphics[width=\textwidth, angle=0,clip=true, scale=1.]{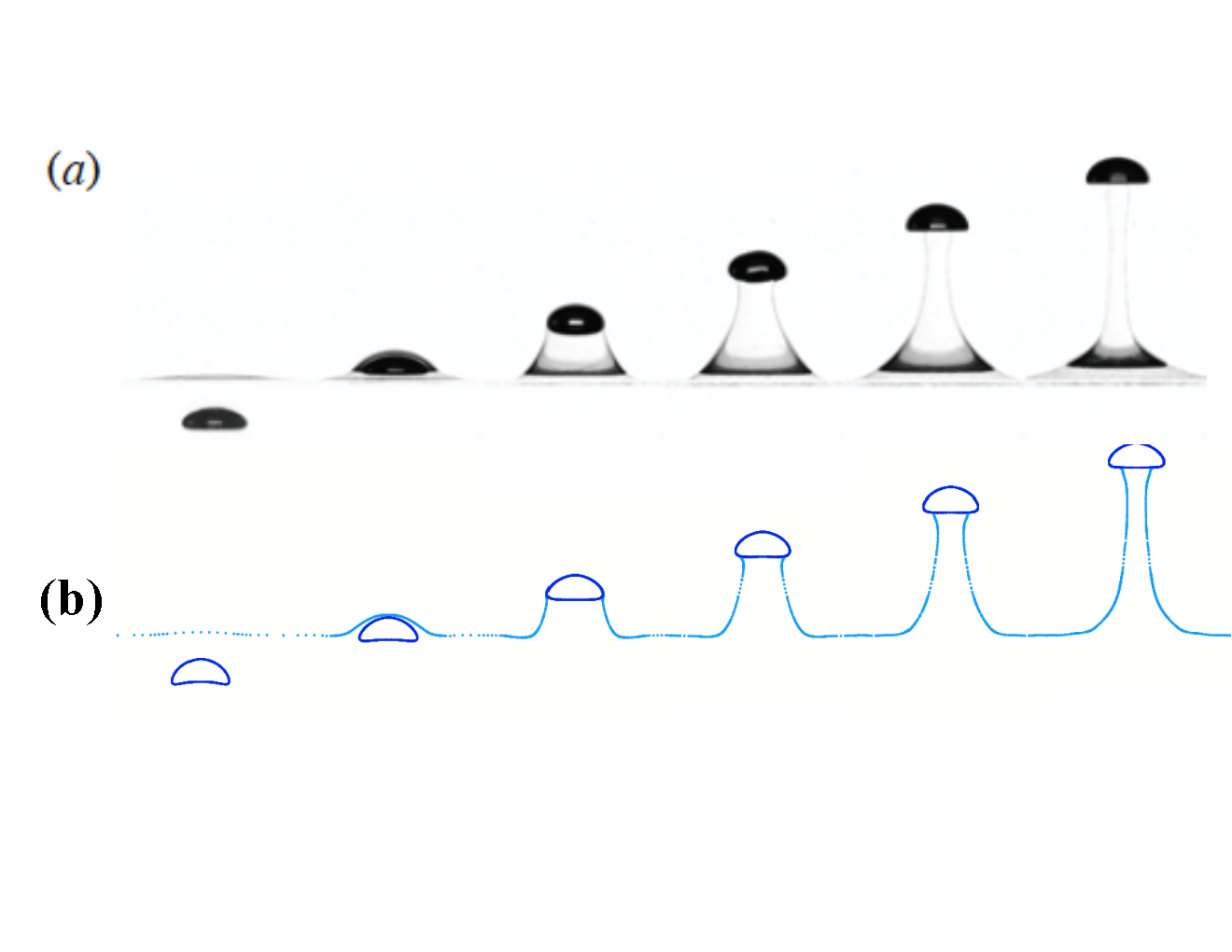}
\vspace{-0.2\textwidth}
\caption{Evolution of shape of a bubble and interface between two fluids as the bubble crosses the interface.  $ Ar=226.8,~Bo=7.375,~Y=0,~\rho=0.794,~m=1.11 $. Here $ \frac{\sigma_{23}}{\sigma_{12}} $ is 0.425 and $ \frac{\sigma_{13}}{\sigma_{12}}$ is 0.615. Figures show (a) the experimental observation  \citep{bonhomme2012inertial}  and (b) our computational results. As the bubble crosses the fluid-fluid interface, entrains a column of heavy fluid into the upper layer. }
\label{fig:Benchmark2}
\end{figure}

We have validated the code by benchmarking our results with the experimental results obtained by \citet{bonhomme2012inertial}. The experiments were based on a single gas bubble crossing an interface between a lower phase made of water or water plus glycerin and an upper, slightly lighter, phase made of silicon oil. Dimensionless numbers from \citet{bonhomme2012inertial} $(Ar_B, Bo_B, \Lambda, R $) are translated to the ones used in this study as follows: : $(Ar=Ar_B^2/8, Bo=Bo_B/4, m=\Lambda, \rho=1-R)$. The interfacial tensions have been transformed to phase specific surface tension using the following relations  \citep{tofighi2013numerical}:

\[
\sigma_1 = 0.5(\sigma_{12}+\sigma_{13}-\sigma_{23}) $$
$$ \sigma_2 = 0.5(\sigma_{12}+\sigma_{23}-\sigma_{13}) $$
$$ \sigma_3 = 0.5(\sigma_{13}+\sigma_{23}-\sigma_{12})
\]

where $ \sigma_{ij} $ shows the surface tension between phase $ i, $ and $ j $ and $\sigma_i$ specifies the surfaces tension of phase $i$. Indices 1, 2, and 3 denote lower layer liquid, bubble, and upper layer liquid, respectively.
The results are plotted in Fig. \ref{fig:Benchmark2}.b against those obtained by \citet{bonhomme2012inertial} in Fig. \ref{fig:Benchmark2}.a. The bubble crosses the interface and tows a column of heavy fluid behind it. The shape of bubble before and after interface, and the column of heavy fluid have been found to compare well with the experimental results.

\subsection{Scope of study}

Our study is based on 167 simulations covering a wide range of governing parameters. The values of governing parameters are selected from a wide range: $ (Y, Bo, \rho, m)= $ \\
$([0, 0.1, 0.15], [0.1, 1, 5, 10, 50], [0.7, 1], [0.1, 1, 10]) $ and various  $ Ar=1, 5, 10, 50, 500$. 

\begin{figure}
\centering
\includegraphics[scale=0.5]{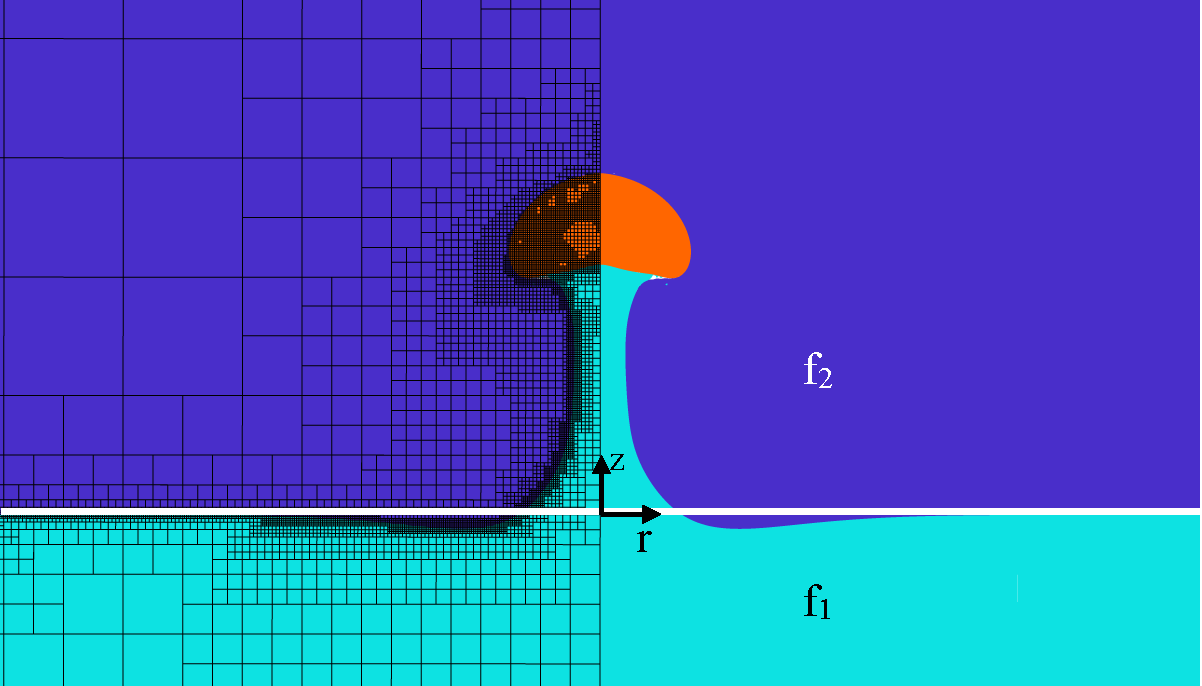}
%\vspace{-2.5cm}
\caption{As the bubble crosses the liquid-liquid interface (between $f_1 $ and $f_2 $), it displaces the initially horizontal interface and tows a column of $ f_1 $. The mesh has been adapted and refined at the interfaces and inside the bubble. The entrainment volume is essentially the cumulative volume of cells $ z>0 $, encompassing $ f_1 $. }
\label{fig:Ve1}
\end{figure}

To quantify the entrainment, the volume of liquid from the lower layer ($ f_1 $) that has been transported to the upper layer ($ f_2 $) is calculated, see Fig.~\ref{fig:Ve1}. As the bubble crosses the liquid-liquid interface, it displaces the initially horizontal interface between the liquid layers and it carries liquid from the lower layer to the upper layer either as a towed column or trapped in the recirculating wake. In this context, we have estimated the entrainment volume by 
\begin{equation}
V_e = \frac{2\pi \Sigma_{z>0} \hat r_{ij} f_1 \hat{\Delta}_{ij}^2C_{ij}}{4/3\pi\hat{R}_b^3} 
\end{equation}

where $ C_{ij} $ is the volume fraction of $ f_1 $, $ \hat{\Delta} $ represents the size of cells at $ z>0 $, encompassing $ f_1 $.This value has been normalized with the volume of rising bubble.

\section{Parametric effects}

In this section we present our results of examining the impact of the governing dimensionless parameters on the entrainment process. 

\subsection{Effect of bubble's shape}\label{sec:shapeR}
 
\begin{figure}
\centering
\includegraphics[scale=0.5]{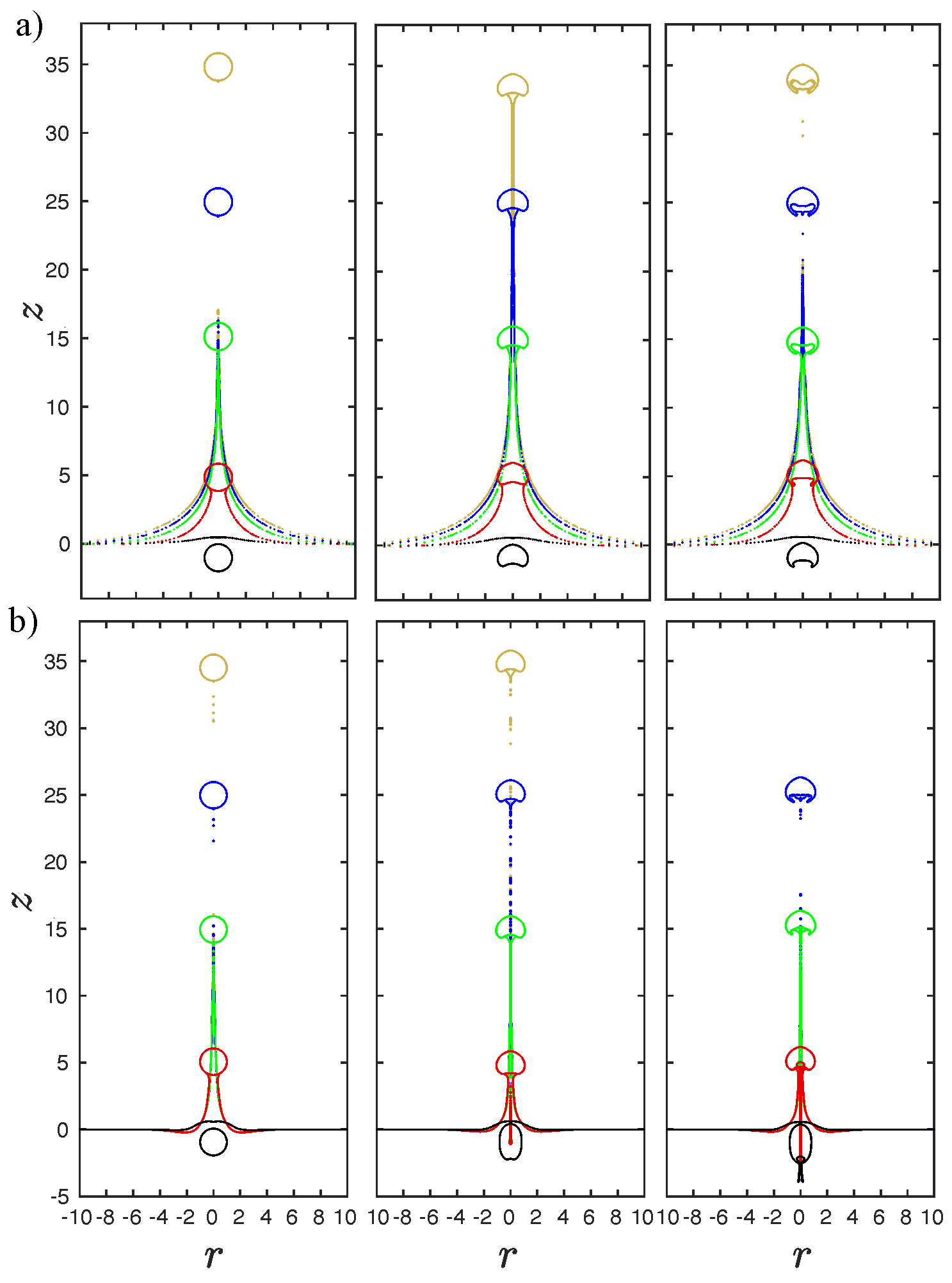}
\caption{Effect of variation of bubble's initial shape on the entrainment at $Ar=5, \rho= 1, m= 1$, for a) $ Y=0 $, b) $ Y=0.1 $ from left to right $Bo=0.1, 10, 50$. The bubble and the evolution of the liquid-liquid interface as bubble centroid reaches five specified heights, $z= -1, {\color{red} 5}, {\color{green} 15}, {\color{blue} 25}, {\color[rgb]{0.8, 0.7, 0.3} 35} $ with black, red, green, and brown colors, respectively, is shown in each panel.}
\label{fig:ShapeR}
\end{figure}

It is of interest to examine how the change in the bubble shape influences the entrainment. We consider two scenarios, (a) both upper and lower liquid layers are Newtonian (b) the lower layer is a yield stress (Bingham) fluid. We begin with a relatively low inertia regime $ (Ar=5) $, where the liquid layers are iso-dense and iso-viscous i.e., $ \rho=1,~m=1 $, and we study the effect of changing $ (Bo) $, Fig.~\ref{fig:ShapeR}.

When the liquid layers are Newtonian, given values of $ \rho$ and $m $, the bubble does indeed rise within a single liquid. The bubble reaches its steady-state shape before reaching the interface between the two liquids and maintains this shape after crossing the interface. By increasing $ Bo $ from 0.1 to 10, the shape of the bubble undergoes a transformation from a spherical shape to a ellipsoidal cap, with wider profile. Tips also develop at the sides of the bubble and the rear side becomes indented. At $ Bo=50 $, a skirted bubble with a smooth and steady skirt forms. 

As depicted in Fig.~\ref{fig:ShapeR} (a), the entrainment behavior varies accordingly. At the beginning, the spherical bubble pulls a portion of the lower layer along with it. After traveling a distance of $z_b \approx \sim 15 $, it eventually separates from the interface. Therefore the main contributing factor to entrainment (at $ Bo=0.1 $) is the displacement of the interface. When the bubble adopts a cap-shaped configuration, its broader profile leads to a more significant displacement of the fluid at the interface compared to when it is spherical. Consequently, a column of liquid from the lower layer accumulates behind the rising bubble. In the case of the indented cap-shaped bubble ($Bo=10$), this liquid column extends beyond $z=35$. The detachment of the bubble from this towed liquid column occurs later, and no entrained liquid remains attached to the bubble. When $Bo$ increases to 50, apart from the interface displacement, the bubble also encapsulates liquid from the lower layer, constituting approximately 20\% of the bubble's volume, forming a new compound. This compound then separates from the column and continues its ascent. At high $ Bo $, similar to low $ Bo $ values, no liquid column forms.

The results indicate that all bubbles rising in a Newtonian liquid cause displacement of the interface. Cap-shaped bubbles, in particular, not only induce interface displacement but also tow behind a liquid column. The thickness of the liquid column is influenced by the width of the bubbles and, eventually, the bubble detaches from the column. On the other hand, skirted bubbles rise as a separate entity, carrying along a portion of the liquid from the lower layer. 

\begin{figure}
\centering

\includegraphics[scale=0.42]{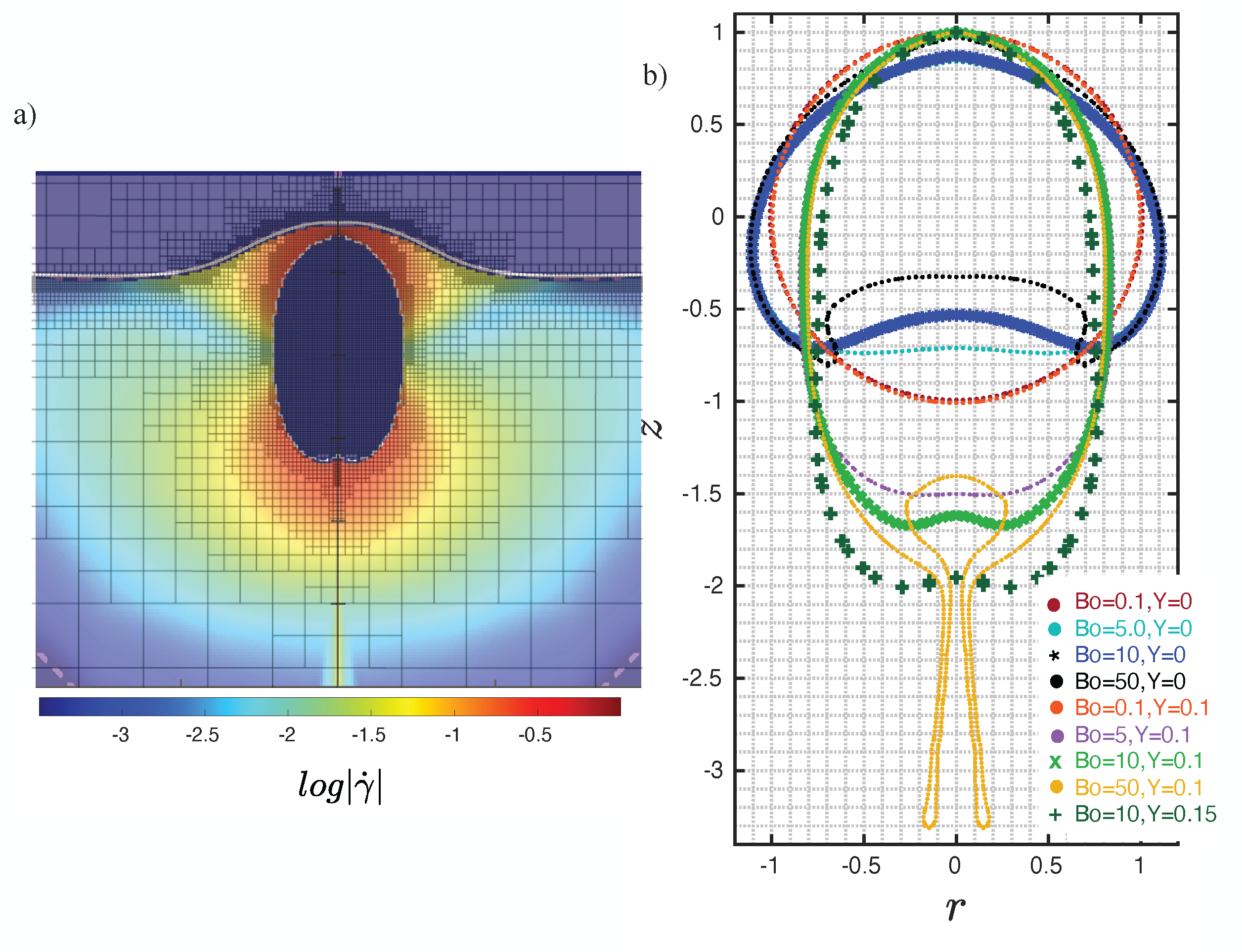}
\caption{(a) Contours of strain rate on the computational cells as a bubble reaches the interface at $Ar=5, Bo=10, Y=0.1, \rho=1, m=1$. (b) The profile of bubble for $Ar=5, \rho=1, m=1$ at different values of $Bo$, including 0.1, 5, 10, and 50, as well as for $Y=0$,  $Y=0.1$ and $Y=0.15$.}
\label{fig:ShapeONve}
\end{figure}

We now turn our attention to the entrainment process when the lower layer is a Bingham fluid. At $ Y = 0.1 $, corresponding to $ Bo $ value, the bubble develops different steady state shapes before reaching the interface between two fluid layers. The steady state bubble shapes varies from a spherical shape to a prolate shape, as $ Bo $ increases. For small values of $Bo \leqslant 1$, the bubble maintains an almost spherical shape, primarily due to the significant influence of surface tension. However, at medium values of $Bo$, the bubble forms a prolate shape. In this case, gravitational forces surpass surface tension effects, resulting in an decreased rate of strain, subsequently increased effective viscosity, around the equatorial plane of the bubble compared to the poles. Consequently, the bubble begins to deform preferentially in the direction of its poles, taking a prolate shape. As $ Bo $ increases and the influence of surface tension decreases, the prolate bubble becomes more elongated. For large $ Bo=50 $, the bubble retains an indentation. These bubble shapes are in agreement with those predicted earlier by \citep{tsamopoulos2008steady}. 

Remarkably, as the bubble approaches the interface, it undergoes increased elongation, as depicted in Fig~\ref{fig:ShapeONve}.a. This elongation is primarily driven by an elevated rate of strain occurring near the north pole where the bubble intersects the interface. This elevated strain rate leads to a reduction in the effective viscosity near the pole, and decreased rate of strain around the equatorial plane, subsequently causing further elongation of the bubble. This elongation becomes even more pronounced with higher values of $Bo$ and $ Y $, as illustrated in Figure~\ref{fig:ShapeONve}.b. We observe two different types of bubble breakup at $Bo=10$ and $Bo=50$. At a moderate value of $Bo$, the bubble experiences breakup as it enters the upper Newtonian fluid, whereas at a high $Bo$, the bubble undergoes breakup within the yield stress layer, with a small portion of it remaining confined within this layer. Both breakup phenomena occur when viscous shear forces surpass the capillary forces, indicating presence of a critical capillary number ($ Ca = \frac{Bo}{Ar} $). The initial breakup of rising bubbles in the yield stress fluid happens at $ Ca_c \approx 5 $ (over the set of our computational parameters). Breakup of drops has been seen at the tip when $m_b = \hat{\mu}_b/\hat{\mu}_p \ll 1$ and for cases with capillary number above a $ Ca_c $ \citep{chu2019review}.

 The contribution of entrainment through the displacement of the interface diminishes when there is a yield stress fluid in the lower layer. It is mainly due to the fact that as the bubble approaches the interface, buoyancy yields only a partial region of the lower fluid while the rest remains unyielded. It can be seen as the passage of bubble from a narrow yielded column to the upper Newtonian layer. Therefore, a limited radial length of approximately 5 of the interface becomes displaced and distances from $ z=0 $. For $ Bo =10 $ and $ 50 $, some Bingham fluid become trapped behind the rising bubble. At $ Bo=10 $, an inverted cone of yield stress fluid remains attached to the bubble and forms a new configuration. The rising bubble and attached cone break off from the towed liquid column around $ z=15 $, while the bubble remains connected to the corresponding Newtonian towed column. In general, the rising bubble transports less fluid to the upper layer and entrainment decreases significantly. 

\begin{figure}
\hspace{-1cm}
\centering
\includegraphics[scale=0.4]{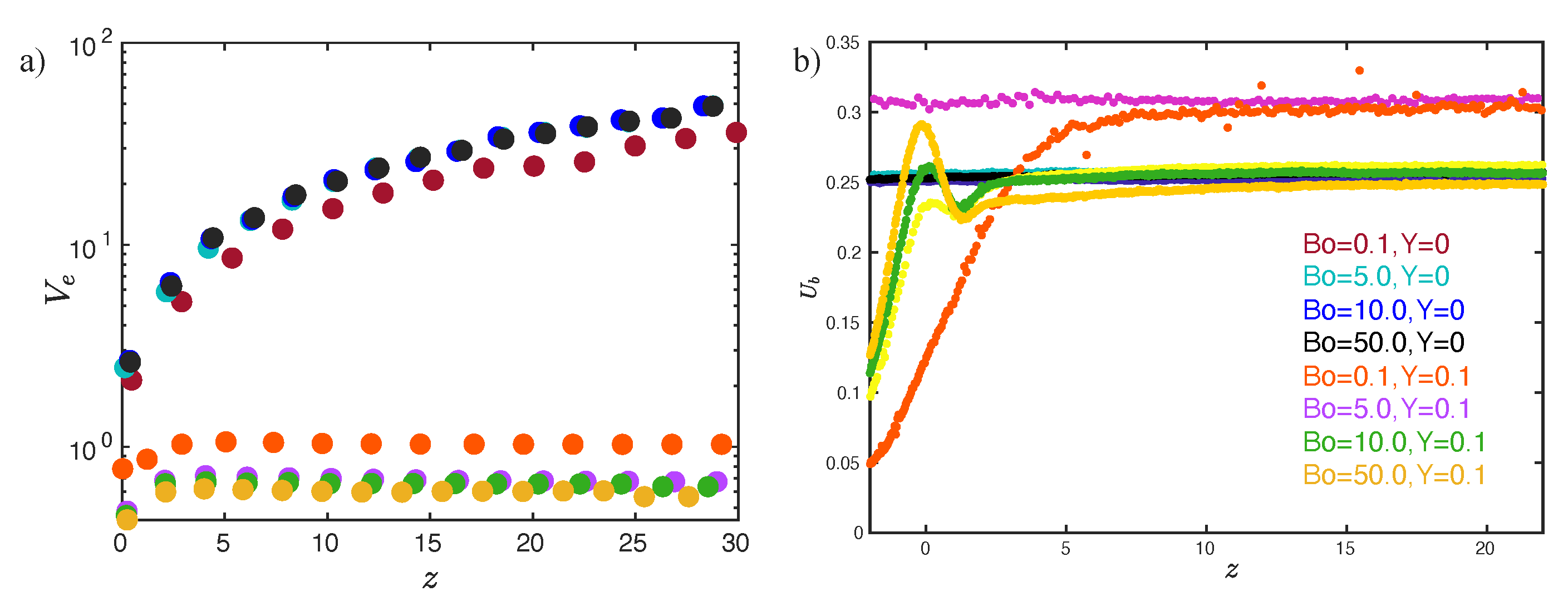}
\caption{Evolution of (a) the volume of entrainment $ V_e $, and (b) the rise speed ($ U_b $) of bubble as a function of the dimensionless bubble position for the cases considered in Fig.~\ref{fig:ShapeR}.}
\label{fig:VeUbAr5}
\end{figure}

The variation of the volume of entrainment as the bubble rises, as well as the rise speed of the bubble for both scenarios (i.e. Newtonian and Bingham lower layers) are shown in Fig.~\ref{fig:VeUbAr5}. The entrainment volume ($V_e$) versus the bubble's position, as shown in Fig.~\ref{fig:VeUbAr5}(a), demonstrates that in the case of Newtonian lower layers, spherical-shaped bubbles transport a smaller amount of liquid to the upper layer when compared to cap-shaped or skirted bubbles. This can be attributed to two factors: (i) The displacement at the interface increases proportionally with the area of the bubble. (ii) The towed column becomes thicker for the same reason. As a result, cap-shaped or skirted bubbles are more efficient in bringing liquid to the upper layer. In contrast, for Bingham lower layers, spherical-shaped bubbles transport a larger quantity of liquid to the upper layer compared to cap-shaped or skirted bubbles. This is because, as $ Bo $ increases, the width of the bubble's profile decreases in the lower yield stress fluid, leading to a thin entrained column and a reduction in entrainment. 

Surprisingly, we do not see a noticable difference in the entrainment for $Bo=5,~10,~50$. By comparing the shape of bubbles and their rising veolocity, one might notice that despite the variety in the shapes, the width of bubbles and their rising velocity are quite similar, see Fig.~\ref{fig:VeUbAr5}(b). It implies that the effect of entrainment on drag forces imposed on the bubble is minimal. Despite that, the rising velocity of the spherical bubble at $ Y=0.1 $ increases gradually and then reaches a steady state value similar to the Newtonian counterpart Fig.~\ref{fig:VeUbAr5}(b). However, the rising velocity at $ Bo=10,~ 50 $ undergoes an acceleration followed by a deceleration. The speed of rising bubble accelerates as it enters the Newtonian upper layer, but due to attached Bingham fluid it undergoes a decceleration.

\subsection{Effect of inetria}\label{sec:inertiaEffect}

We investigated the effect of inertia on the transport of both Newtonian and yield stress fluids at moderate and high values of Bond numbers, $Bo=5$ and $ Bo=50 $. 

\begin{figure}
\centering
\includegraphics[scale=0.5]{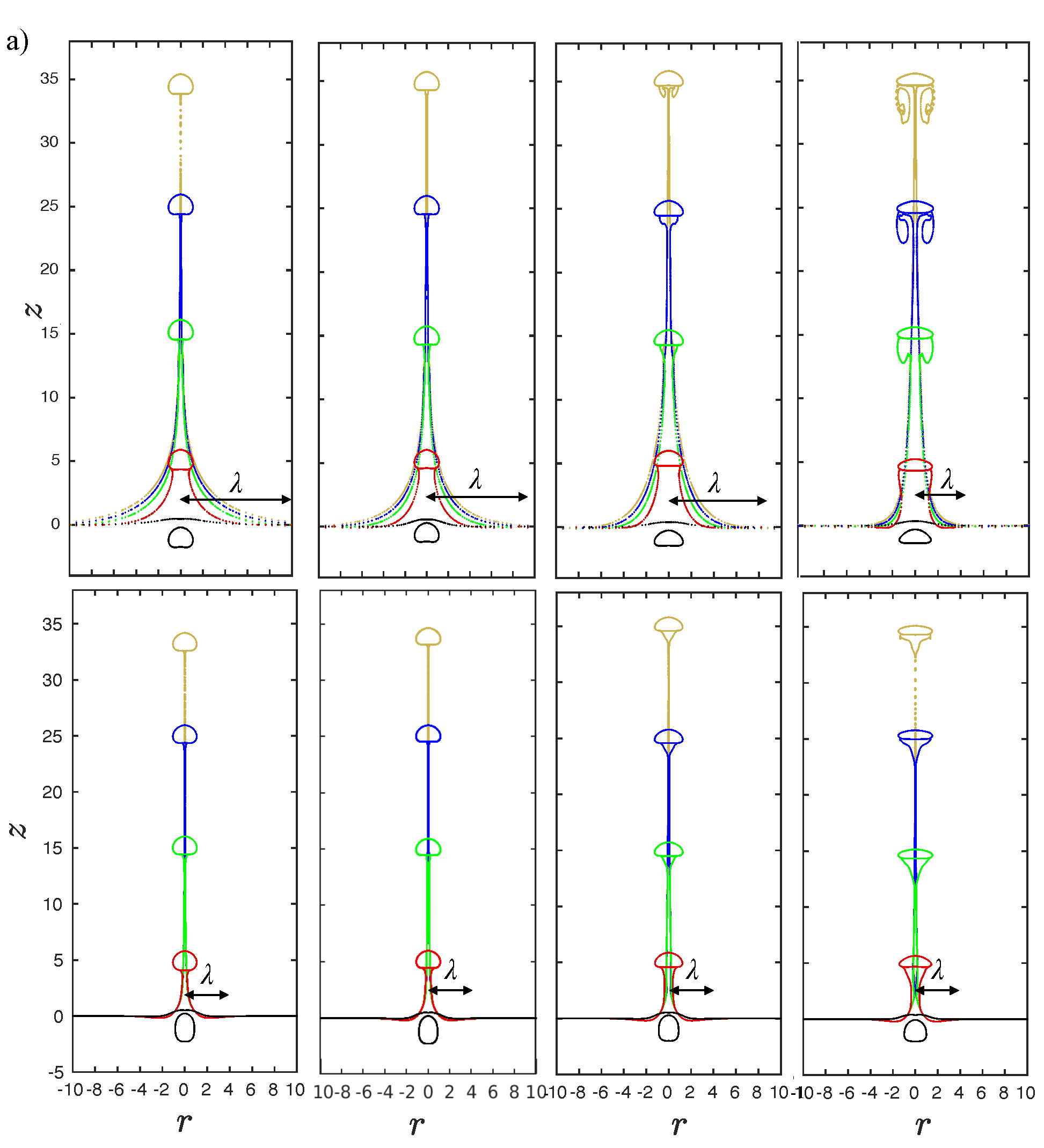}
\caption{Effect of increasing Ar on the entrainment at moderate Bo; $Bo=5$ and $\rho=1, m=1$ for a) $ Y=0 $, b) $ Y=0.1 $ from left to right $Ar=5,10,50, 500$. For colors refer to the caption of Fig.~\ref{fig:ShapeR}.}
\label{fig:ArR5}
\end{figure}

The Newtonian results for $Bo=5$ are presented in Fig.~\ref{fig:ArR5}(a). As we discussed in Sec. \ref{sec:shapeR}, as the bubble approaches the initially horizontal Newtonian-Newtonian interface, the interface becomes displaced. This displacement occurs within a specific radial domain and dissipates at a critical radius denoted $\lambda$, see Fig.~ \ref{fig:ShapeR}. By increasing $ Ar $, $ \lambda $ decreases $ \sim 10,~9.7,~8,~4$. $ Ar $ can be interpreted as a Reynolds number, with characteristic velocity $ \hat U $ from the balance of buoyant and viscous stresses. As $ Ar$ (i.e. $ \hat U$) increases the ratio of advective to vsicous timescales decreases and disturbance of the interface behind the bubble (i.e. $ \lambda $) decreases.

 The bubble develops different shapes before reaching the interface. It undergoes a transition from a spherical-cap to an ellipsoidal cap with a wider profile as $Ar$ increases. This alteration in shape leads to a wider towed column. Subsequently, the length of the towed column extends as $Ar$ increases. For $ Ar=5,~10 $, no liquid remains attached to the bubble and the bubble leaves off the towed column before reaching top of the computational domain ($ z= 50$). At $Ar=50$, some liquid remains initially attached to the bubble; however, it gradually lags behind and ultimately separates from the bubble. In the highly inertial regime, specifically at $Ar=500$, the entrained liquid divides into two parts: a towed column and a recirculating torus behind the bubble. The towed column becomes progressively thinner, and the recirculating bodies gradually move further away from the bubble. 

The yield stress results for $Y=0.1$ are depicted in Fig.~\ref{fig:ArR5}(b). In this scenario, there is no significant change in $\lambda$ with the increase in $Ar$, and only a small portion of the interface moves away from $z=0$. The interface remains nearly flat as the bubble crosses it and ascends further away. The shape of bubble before reaching the interface varies by $ Ar $. As $Ar$ increases, the aspect ratio decreases, causing the profile to shift from a prolate shape to one more akin to an oblate shape with a flat rear side. The main contribution to the entrainment comes from the Bingham fluid attached to the rising bubble at $ Ar=50~\&~500$. In contrast with the Newtonian counterpart, the inverted cone of yield stress fluid remains attached to the bubble and doesn't decay with time/distance. The fluid in the inverted cone is partly unyielded. At $Ar=500$, the rate of strain within the Newtonian fluid surrounding the towed column is higher than that of Bingham fluid inside the column. Consequently, the interface becomes concave, leading to a significantly different flow pattern around the bubble and entrainment, compared to the Newtonian counterpart.

\begin{figure}
\centering
\includegraphics[scale=0.5]{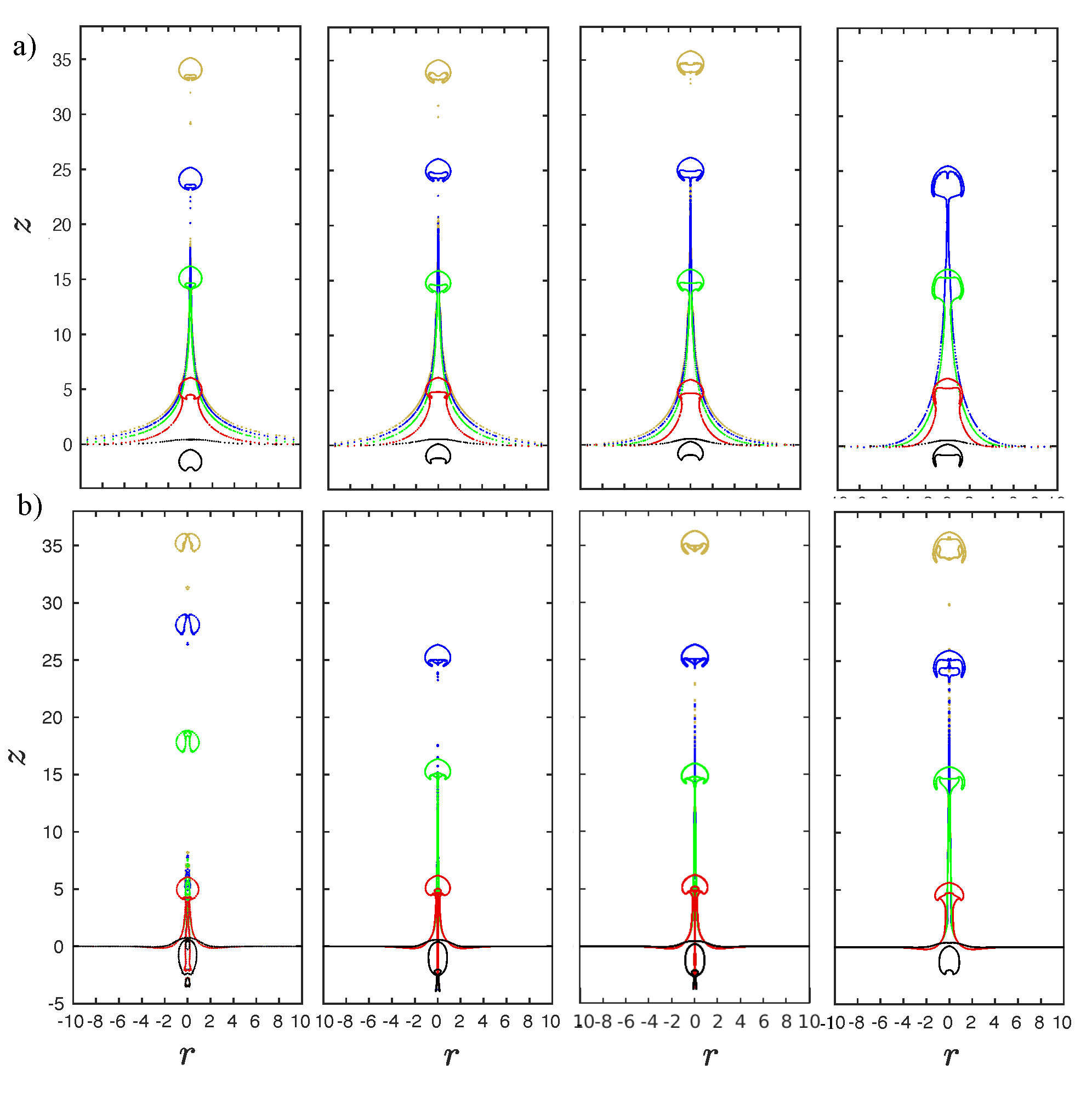}
\caption{Effect of increasing Ar on the entrainment at large Bo; $Bo=50$ and $\rho=1, m=1$, for a) $ Y=0 $, b) $ Y=0.1 $ from left to right $Ar=1,5,10,50$. For colors refer to the caption of Fig.~\ref{fig:ShapeR}. }
\label{fig:ArR50}
\end{figure}

The results for $ Bo=50 $, and $ Ar=1, 5, 10, 50 $ are shown in Fig.~\ref{fig:ArR50}. When the lower layer is a Newtonian fluid, by increasing $ Ar $, the profile of bubble becomes wider and unstable skirts form. As a result, more liquid becomes trapped behind the bubble and larger ascending compounds form. At $Ar=50$ and $Bo=50$, the bubble becomes highly unstable and splits into two parts at $z \sim 25$. 

The impact of inertia on the displacement of a yield stress fluid ($Y=0.1$) is depicted in Fig.~\ref{fig:ArR50} (b). As $Ar$ increases, the bubble becomes wider and consequently, the thickness of the entrained column increases. At low and medium values of $ Ar $, where $ Ca \geqslant 5 $, bubble breaks up before entering the upper layer and some air becomes entraped within the yield stress fluid. The increase in $ Ar $, causes a greater amount of liquid to trap behind the ascending bubble wihtin the upper Newtonian layer.

\subsection*{Upper and Lower layers with similar properties}

So far our results show how a rising bubble induces flow and transports liquid at specific values of $ Ar$ and $Bo $. As discussed in Secs.\ref{sec:shapeR} and \ref{sec:inertiaEffect}, the lower layer liquid could be transported by either the displacement of the interface (in the case of spherical shaped bubbles), or the towed column (in the case of cap-shaped bubbles) or the wake behind a rising (skirted shaped) bubble. When bouyancy forces are dominant, high $ Bo $ regime, the liquid attached to the cap shaped bubbles break off from the column and rises up with the bubble. Increasing inertia in each regime amplifies the process and increases the retention time at low and moderate $ Bo $ values. 

\subsection{Effect of viscosity}

\begin{figure}
\centering
\includegraphics[scale=0.5]{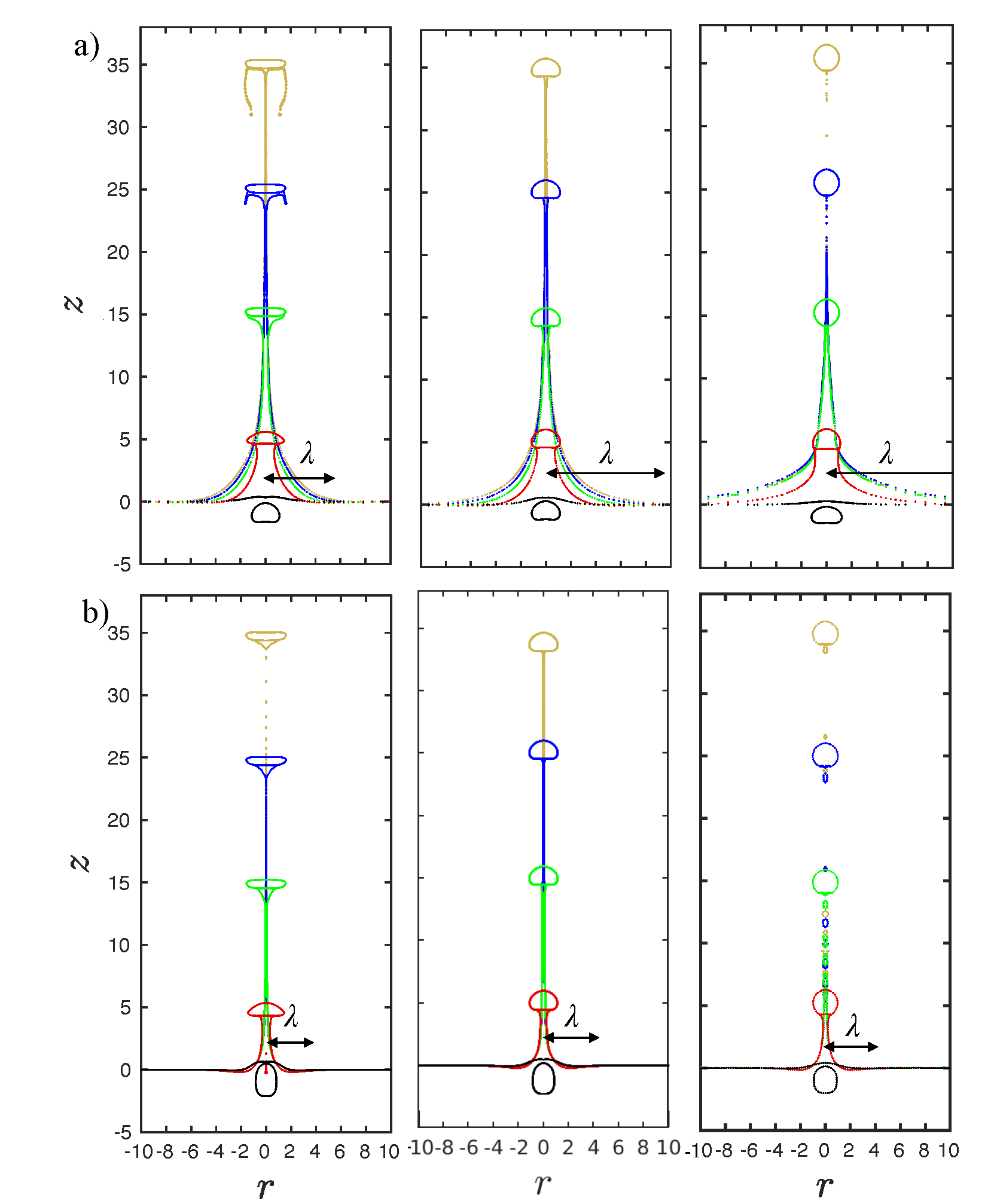}
\caption{Effect of variation of the viscosity ratio ($ m $) on the entrainment for $Ar= 10, Bo=5, \rho=1$ and (a) $ Y=0 $, (b) $ Y=0.1 $; $ m $ increases from left to right as $ 0.1,1,10 $. For colors refer to the caption of Fig.~\ref{fig:ShapeR}. }
\label{fig:mucapshaped}
\end{figure}

 When liquid layers have different viscosities but similar densities, the viscosity ratio ($m$) plays a significant role in altering the Archimedes number in the upper layer. An increase in $ m $ results in a decrease of $Ar$ by $ \frac{1}{m^2} $ while the Bond number remains unaffected. We investigated the effect of varying $ m $ on the rise of bubble and entrainment in moderate $Ar$ and $ Bo $ regime, where bubbles form an indented cap shape. The passage of the bubbles and subsequent entrainment at $ m=0.1,~1,~10 $ are shown in Fig. \ref{fig:mucapshaped}.

 The results for the case in which the upper and lower layers are Newtonian are shown in Fig.~\ref{fig:mucapshaped}(a). By increasing $ m $, $ \lambda $ increases $ \sim 8, 14, 46  $, which is counterintuitive. By increasing $ m $, the bubble experiences deceleration and consequently the advective to viscous time scale increases and the disturbance of the interface behind the bubble (i.e. $ \lambda $) increases.
% 
%  {\color{red} It can be explained by the fact that the energy that created the displacement of the interface comes from the rising bubble, and as time goes, this energy gets spread over a larger and larger circumference. When $m=10$, the bubble experiences deceleration and consequently requires more time to cross the interface in comparison to the case when $m=0.1$. Hence, $ \lambda $ increases. However, one might argue that the rising velocity of the bubble at the interface is not the same for these two cases also the bubble's energy at the interface, what you think?  } 

When the bubble enters the upper layer at $ m=0.1 $, the Archimedes in the upper layer increases to 1000, so we expect that the bubble becomes wider and the entrained column becomes thicker. However, contrary to these expectations, the increase in Archimedes (to 1000) does not lead to a proportional increase in the entrainment, and the width of the liquid column does not become significantly thicker with respect to $Ar=10, m=1$. This could be attributed to the high shear stresses that are generated in the upper layer, which cause the towed liquid column to have a thickness comparable to the width of the bubble. On the other hand, when the viscosity ratio is $m=10$, as the bubble enters the upper layer, the Archimedes number decreases to 0.1. In this situation, the bubble's inertia becomes significantly less significant than the viscous stresses. Consequently, the bubble deforms and takes on a spherical shape, as expected. The entrainment in this case is caused by a significant displacemnet of the interface. Due to the initial shape of the bubble and less viscosity of the lower layer, which results in generation of higher strain rates, the bubble displaces a larger area of the interface and creates an initially thick towed column. However, as the flow regime transitions and the shape of the bubble changes, it separates from the Newtonian-Newtonian interface and the liquid column is not extended to $ z>15 $.

When the lower layer is a Bingham fluid, as illustrated in Fig.~\ref{fig:mucapshaped}(b), as the bubble approaches the interface, its aspect ratio increases. This elongation becomes more pronounced when the Newtonian viscosity is less than the plastic viscosity ($ m=0.1 $). In this case, the bubble breaks up when it is near complete emergence from the interface. Formation of break-up bubbles is caused by the acceleration upon entering the upper layer as it has higher Archimedes. These break-up bubbles become trapped between the two liquid layers. In the case of ($ m=10 $), the bubble assumes a more spherical shape upon entering the upper layer. The bubble initially tows a liquid column, and therefore the trailing part of the bubble becomes slightly unstable, giving rise to the formation of small drops. The entrainment of yield stress fluids manifests as the towed liquid column for $ m\leqslant 1 $. When $ m >1 $, the towed column is short and truncated around $ z=5 $. In contrast to Newtonian fluids, there is no significant change in $\lambda  $ by varying $ m $ .

\subsection{Effect of density variation}

\begin{figure}
\centering
\includegraphics[scale=0.5]{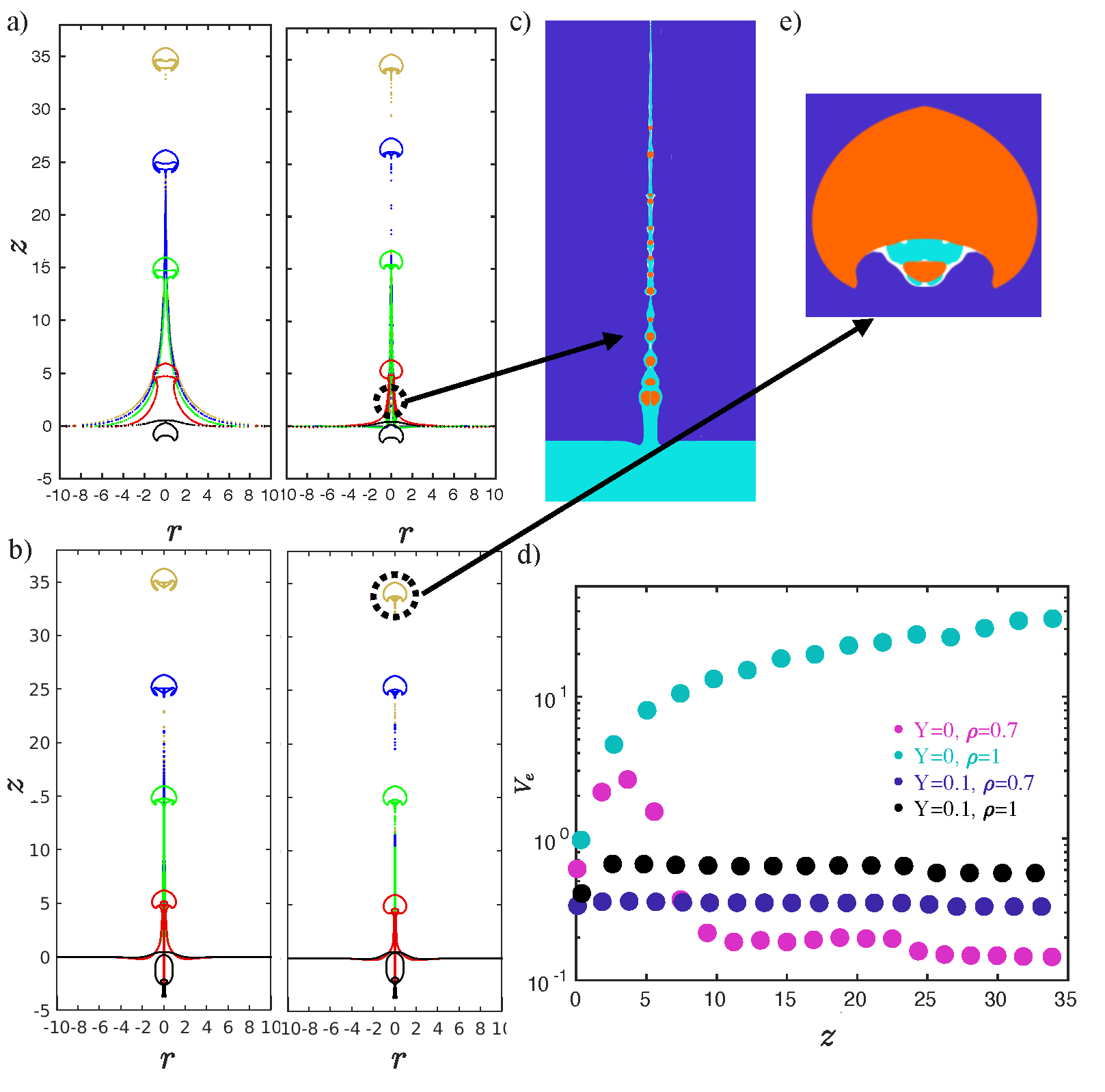}
%\vspace{-2.5cm}
\caption{Effect of density variation on the entrainment for $Ar= 10,~Bo=50,~\&~m=1$: a) $Y=0$; b) $Y=0.1$; The left figures show the entrainment for $\rho=1$, and the right figures show the entrainment when $\rho=0.7$. For colors refer to the caption of Fig.~\ref{fig:ShapeR}; (c) Fragmentation of bubble into smaller bubbles as it enters the upper layer; (d) The volume of entrainment ($ V_e $) as a function of bubble position; and (e) entrapment of heavy Bingham fluid as well as a small bubble in the wake of rising bubble.}
\label{fig:DensityR}
\end{figure}

By keeping liquid layers iso-viscous and reducing the density of the upper layer to $ 70\% $ of the lower layer, the Archimedes decreases by a factor of 0.49 and the Bond number decreases by a factor of 0.7. To study these effects, we considered $ Bo=50 $, at $ Ar=10 $, for $ Y=0,~\&~Y=0.1 $ the results are shown in Fig.~\ref{fig:DensityR}. 

The reduction in Archimedes and Bond, both suggest the bubble deformation to a more indented spherical cap shape upon entering the upper layer. When the lower layer is Newtonian, this transformation causes the bubble to fragment into smaller bubbles as it departs from the interface, see Fig.~\ref{fig:DensityR}(c). Interestingly, the small bubbles generated above the interface are encapsulated within the towed heavy Newtonian liquid transport some of the surrounding liquid and prevent the towed column from descending. 

The results for the case where both upper and lower layers are Newtonian are displayed in Fig.~\ref{fig:DensityR}(a). As the bubble enters the lighter fluid, the displacment at the interface and the subsequent entrainment noticably decreases. The column of the entrained liquid thins and nearly deminishes as the bubble reaches $ z=5 $. In contrast to the iso-dense liquid layers in which the interface rises until the bubble detaches, here the interface gradually settles back. 

 The yield stress results are shown in Fig.~\ref{fig:DensityR}(b). When the lower layer is a yield stress fluid, the bubble tows a column of this fluid, but the towed liquid doesn't change with time; it remains stationary above the interface. The density difference is not sufficient to cause the entrained liquid to settle back after the bubble breaks off. However, this density difference becomes apparent in the thinned towed column. Interstingly, there is some heavy Bingham fluid trapped in the wake of rising bubble as well as a small bubble, Fig.~\ref{fig:DensityR}(e). The small bubble has been formed during shape transfomation above the liquid-liquid interface,
 
 In both cases i.e. Newtonian and Bingham lower layer, bubbles tow some liquid as a column, which settles down shortly after the bubbles travels a distance of $z \sim 5$, and also entrap some in their wake. Initially, a greater amount of liquid is transported from a Newtonian layer compared to a yield stress fluid layer. However, a portion of the Newtonian entrained liquid settles down due to the density difference. Consequently, the entrained heavy Newtonian liquid is less than the entrained heavy yield stress liquid  see Fig.~\ref{fig:DensityR}(d).

\begin{figure}
\centering
\includegraphics[scale=0.5]{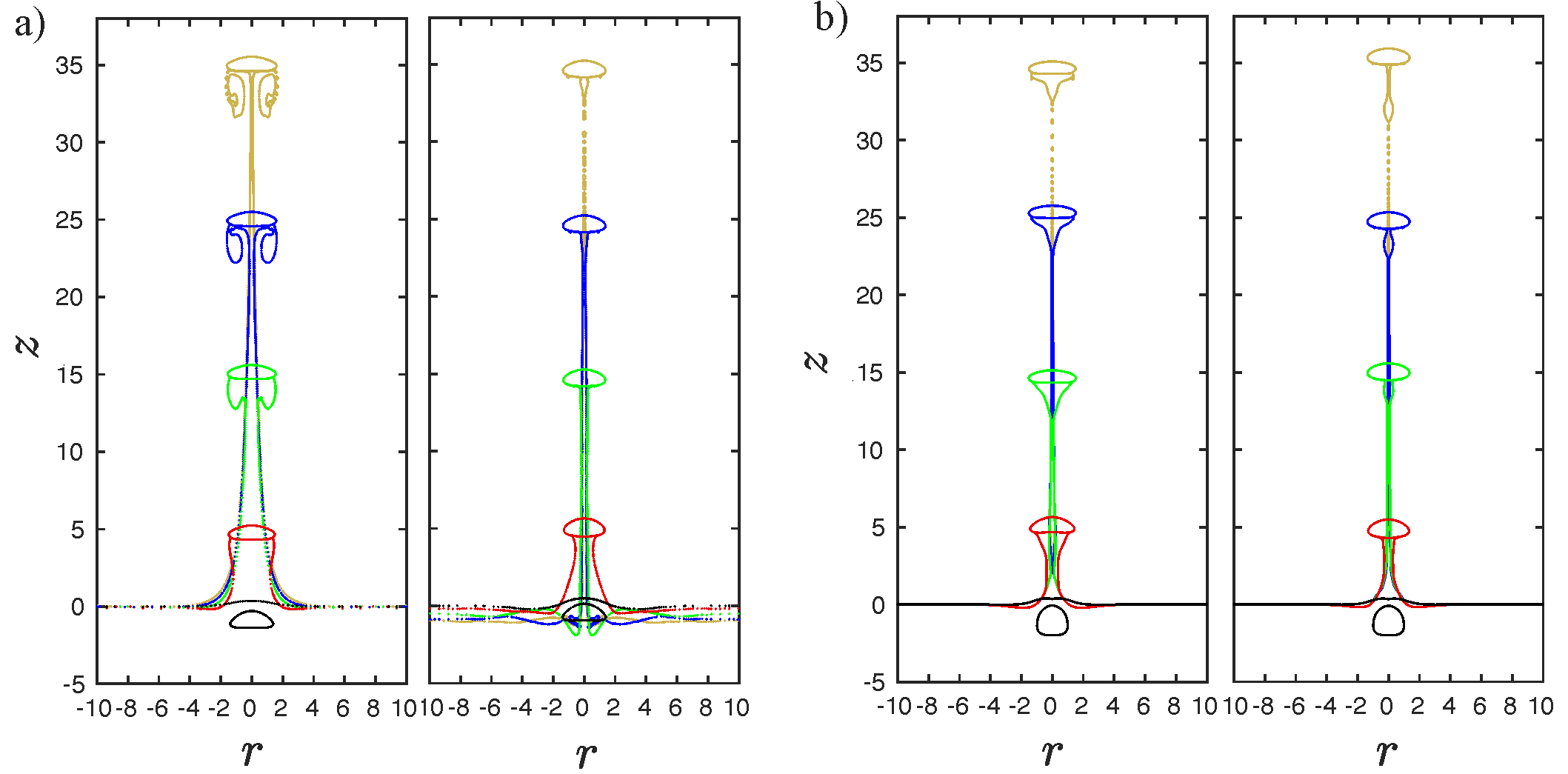}
%\vspace{-2.5cm}
\caption{Effect of density variation on the entrainment for $Ar= 500,~Bo=5,~\&~m=1$; a) $Y=0$; b) $Y=0.1$; The left figures show the entrainment for $\rho=1$, and the right figures show the entrainment when $\rho=0.7$. For colors refer to the caption of Fig.~\ref{fig:ShapeR}. }
\label{fig:DensityR2}
\end{figure}

The results of high-inertia regime with $Ar=500$ are shown in Fig.~\ref{fig:DensityR2}. When the bubble rises from a heavy Newtonian layer to a light Newtonian layer, it assumes a wide topology both below and above the interface. As a result, the bubble doesn't undergo fragmentation as it enters the lighter fluid, and the heavier towed column splashes onto the liquid-liquid interface, see Fig.~\ref{fig:DensityR2}(a). Interestingly, the interface connecting the bubble to the towed column changes from convex to concave. The negative bouyancy force causes the strain rates in the towed column near the equatorial plane around the bubble to become less than those in the surrounding fluid. This results in a change in the cancavity of the interface and absence of the recirculating regions behind the rising bubble. The entrained heavy liquid, visible as the towed column and wake behind the bubble, gradually settle down. 

The results for the case with a Bingham fluid in the lower layer are shown in Fig.~\ref{fig:DensityR2}(b). The main difference between the entrainment at $ \rho=0.7 $ with $ \rho=1 $ is in the entrapment of liquid behind the rising bubble in the upper layer. At $ \rho=1 $, an inverted cone of Bingham fluid forms behind the rising bubble while at $ \rho=0.7 $, the liquid behind separates from the rising bubble and falls down.

\subsection{Effect of density and viscosity variation}

\begin{figure}
\centering
\includegraphics[scale=0.5]{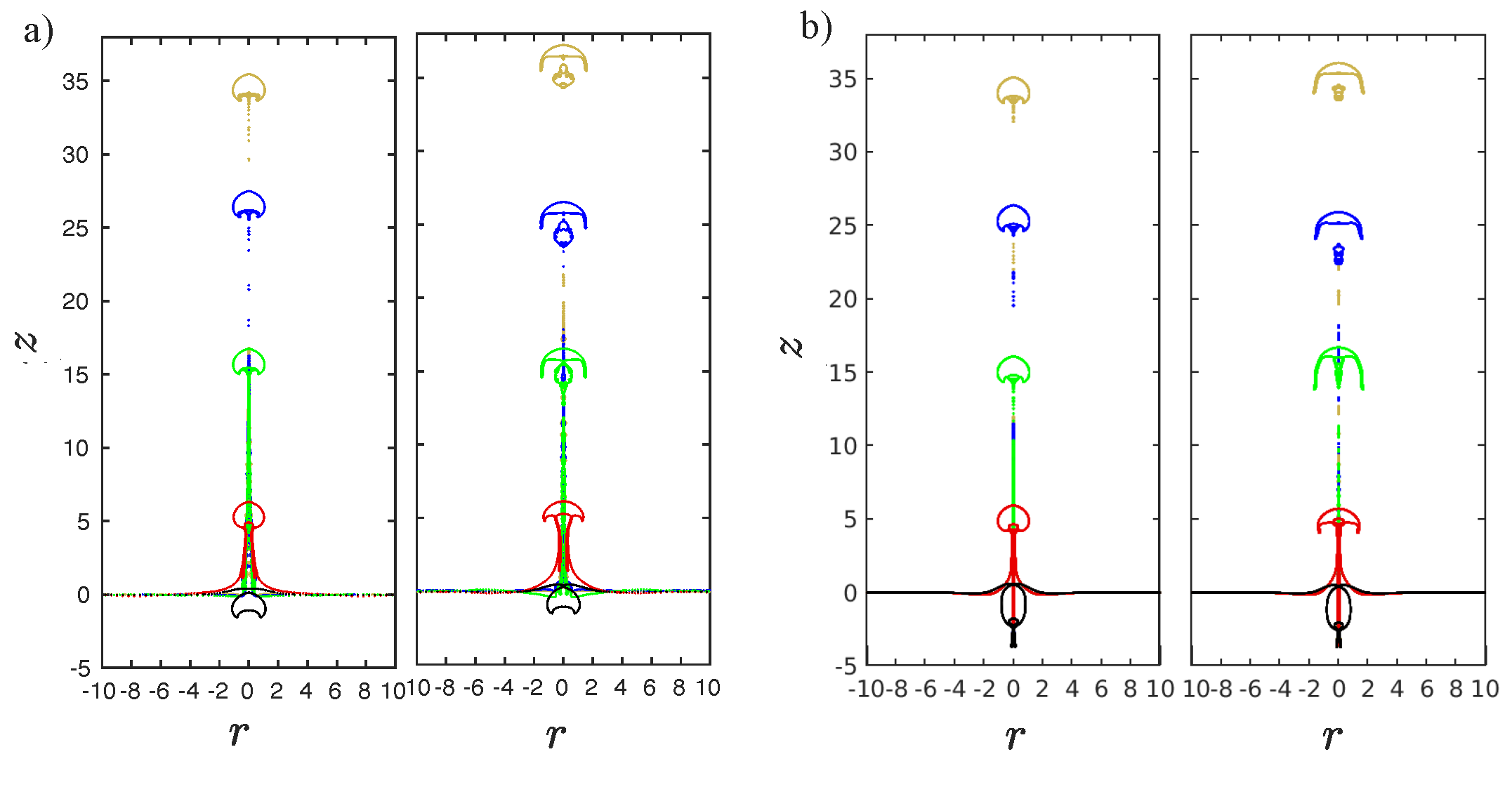}
%\vspace{-2.5cm}
\caption{Effect of density and viscosity variation on the entrainment for $Ar= 10,~Bo=50,~\&~\rho=0.7$; a) $Y=0$; b) $Y=0.1$. The left figures show the entrainment for $m=1$, and the right figures show the entrainment when $m=0.1$. For colors refer to the caption of Fig.~\ref{fig:ShapeR}. }
\label{fig:rhom}
\end{figure}

It is of interest to explore the entrainment in the case that Archimedes number increases as the bubble crosses the interface, but the Bond number decreases. The results obtained for  $ Ar=10, Bo=50, \rho=0.7$ at $m=1~\&~m=0.1$ are shown in Fig.\ref{fig:rhom}.

In both cases of the Newtonian and Bingham lower layers, as the bubble rises and encounters the upper layer, it adopts a wider profile and develops an unstable skirted shape. The bubble's profile accommodates transportation of more liquid to the upper layer. However, notably, the entrained liquid does not remain within the wake of the skirted bubble and instead descends. This presents an interesting contrast with the case of $ \rho=0.7 $ and $ m=1 $. Furthermore, the bubble breaks up in the lower Bingham fluid before reaching the interface, $ Ca=5 $, and a small portion of bubble remains within the lower layer. 

\section{Regimes}

As the bubble crosses the interface, the inteface changes its shape and a portion of lower layer fluid is entrained into the upper layer fluid. During the ascent of the bubble in the upper layer, the entrainment volume changes by time. We have identified the following four different chracteristic behaviours.
 
\begin{enumerate}
\itemindent=11pt
\item[i)]~\textit{Plateau regime} - The rising bubble breaks off from the liquid-liquid interface, either with or without carrying a portion of entrained fluid in its trailing wake. Upon the separation of the bubble from the entrained fluid, it does not return to the lower layer. Consequently the entrained volume reaches a \textit{Plateau} and remains relatively unchanged over time.
\item[ii)]~\textit{Descending - steady state regime} - Initially, the bubble transports some fluid, but the fluid separates from the liquid-liquid interface. It may do so either with or without carrying some fluid in its trailing wake. In either case, a portion of the entrained fluid settles down. Eventually, the entrained volume reaches a plateau and doesn't change over time.
\item[iii)]~\textit{Descending regime} - If the reduction in $ V_e $, explained in (ii), continues it's called \textit{Descending}.
\item[iv)]~\textit{Ascending regime} - The rising bubble remains connected to the liquid-liquid interface as it rises up, so the entrained volume is \textit{Ascending}., i.e. $ V_e $ keeps increasing.
\end{enumerate}

\begin{figure}
\hspace*{-0.1\textwidth}
\centering
\includegraphics[width=\textwidth, angle=0,clip=true, scale=1.2]{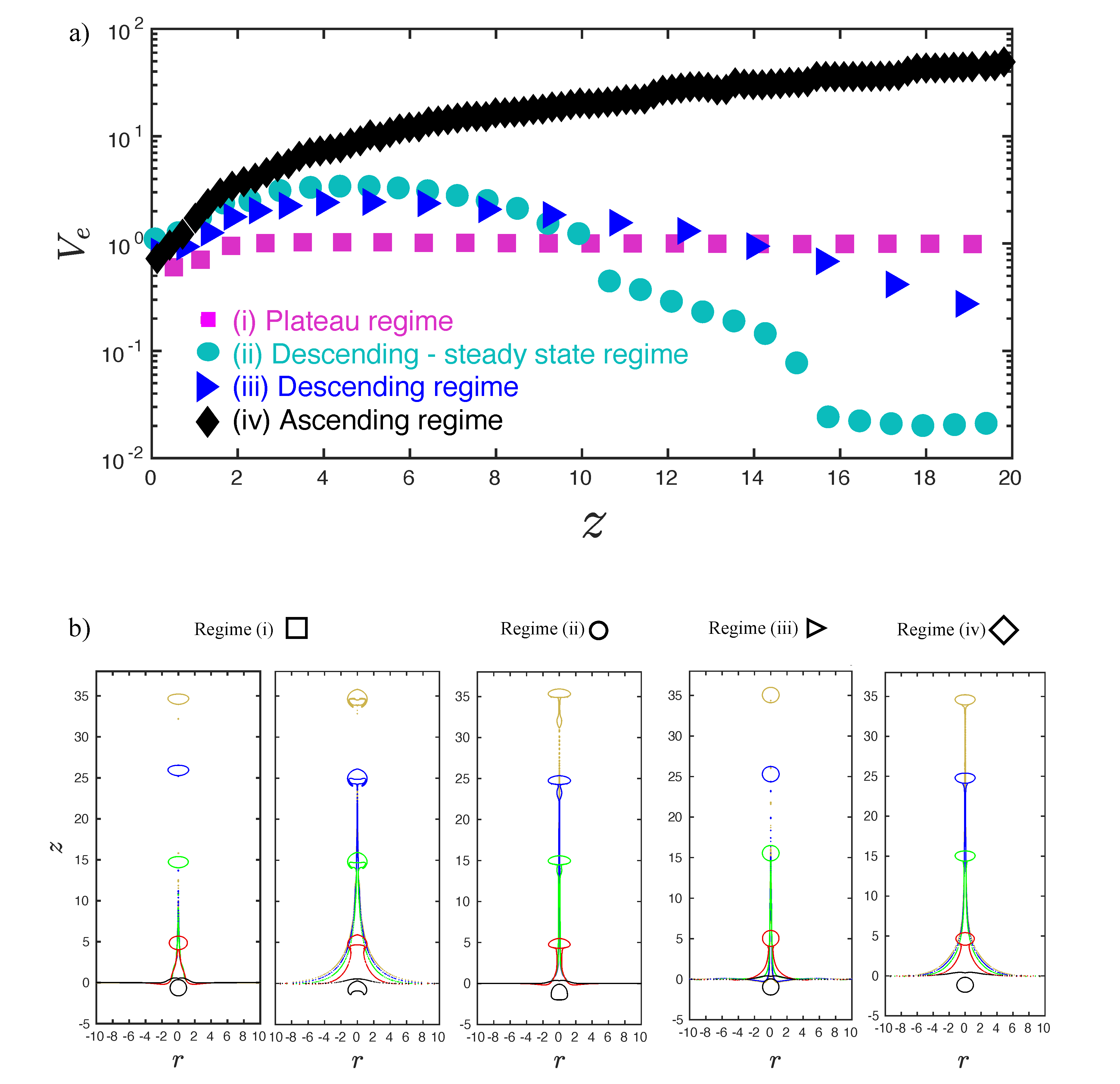}
\caption{ (a) The variation of entrainment with bubble position for four distinct entrainment regimes. (b) Representative examples of four entrainment regimes, for colors refer to the caption of Fig.~\ref{fig:ShapeR}. }
\label{fig:Regimes}
\end{figure}

The variation of entrainment with bubble position for these four distinct entrainment regimes are shown in Fig.~\ref{fig:Regimes}(a) and respresentative examples of the the evolution of the interfaces are shown in Fig.~\ref{fig:Regimes}(b), for each case.

It is of interest to identify key parameters controling the entrainment regimes. The entrainment regimes are investigated for $ Y=0~\&~0.1 $ and $ \rho=1~\&~0.7 $, and are depicted in the domain of $ Ar-Bo $ for different values of $ m $ as follows. 

\begin{figure}
\centering
\includegraphics[width=\textwidth, angle=0,clip=true, scale=.8]{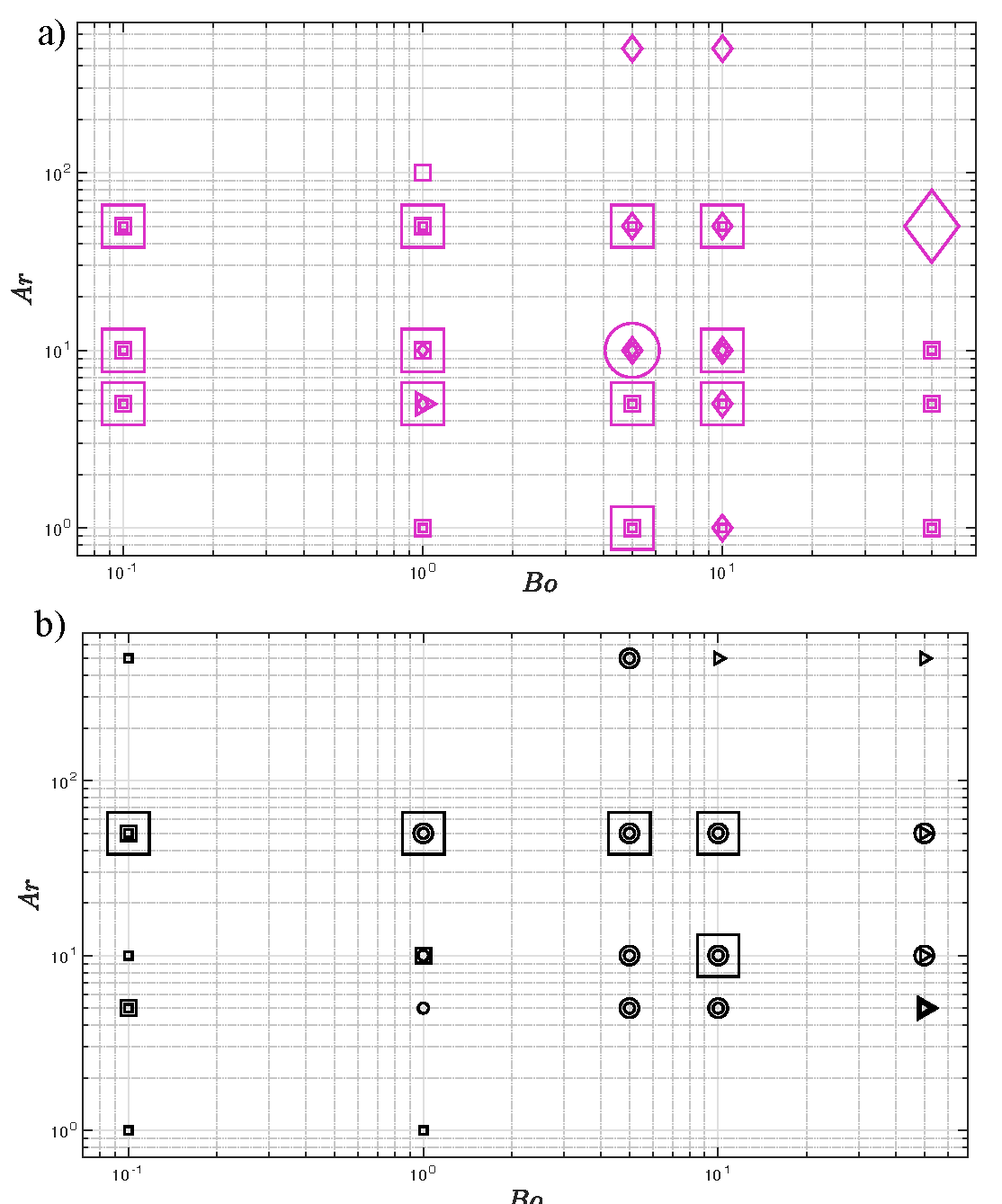}
\caption{Classification of entrainment regimes over the Ar-Bo domain for (a) $ \rho=1 $ and $ Y=0 $ (b) $ \rho=1 $ and $ Y=0.1 $. The size of symbols represents the value of $ m $ at each point, i.e $ m=0.1,~1,~10 $. Regimes are represnted by symbols as follows: $\square$ \textit{Plateau}, $\circ$ \textit{Descending - steady state}, $\triangleright$ \textit{Descending}, and $\diamond$ \textit{Ascending}.}
\label{fig:Regimesrho1}
\end{figure}

The entrainment regimes for $ \rho=1, Y=0 $ are shown in Fig.~\ref{fig:Regimesrho1}(a). For $ \rho=1 $, there is no driving force for entrained fluid to settle. In this case \textit{Plateau} and \textit{Ascending} regimes occur. At low Bo, the dominant regime is the \textit{Plateau} regime. The bubble assumes a spherical shape, and the entrainment happens only in the form of displacement of the interface. At moderate Bo and $ m=0.1~\&~1 $, the gravitational forces are stronger than surface tension, and the bubble remains connected to the liquid-liquid interface and a towed column forms. There are no significant forces leading to break off of bubble with the column. So, we may observe \textit{Ascending} regime in this case. This is the only case where the \textit{Ascending} regime occurs.

The entrainment regimes for $ \rho=1, Y=0.1 $ are shown in Fig.~\ref{fig:Regimesrho1}(b). In this case \textit{Plateau} and \textit{Descending - steady state} regimes occur. At low $ Bo $, the dominant regime is the \textit{Plateau} regime. The bubble assumes a spherical shape, and the entrainment is limited to formation of low towed columns above the interface. These transported yeild stress fluids remain in the upper layer. At moderate to large $ Bo $, two behaviours are observed depending on the value of $ m $: \textit{Descending - steady state} or \textit{Plateau} regime.  If any liquid becomes trapped in the wake behind the rising bubble, the entrainment regime is identified as \textit{Descending - steady state} ($m\leq1$); otherwise, it is denoted as the \textit{Plateau} regime ($m>1$).

\begin{figure}
\centering
\includegraphics[width=\textwidth, angle=0,clip=true, scale=.8]{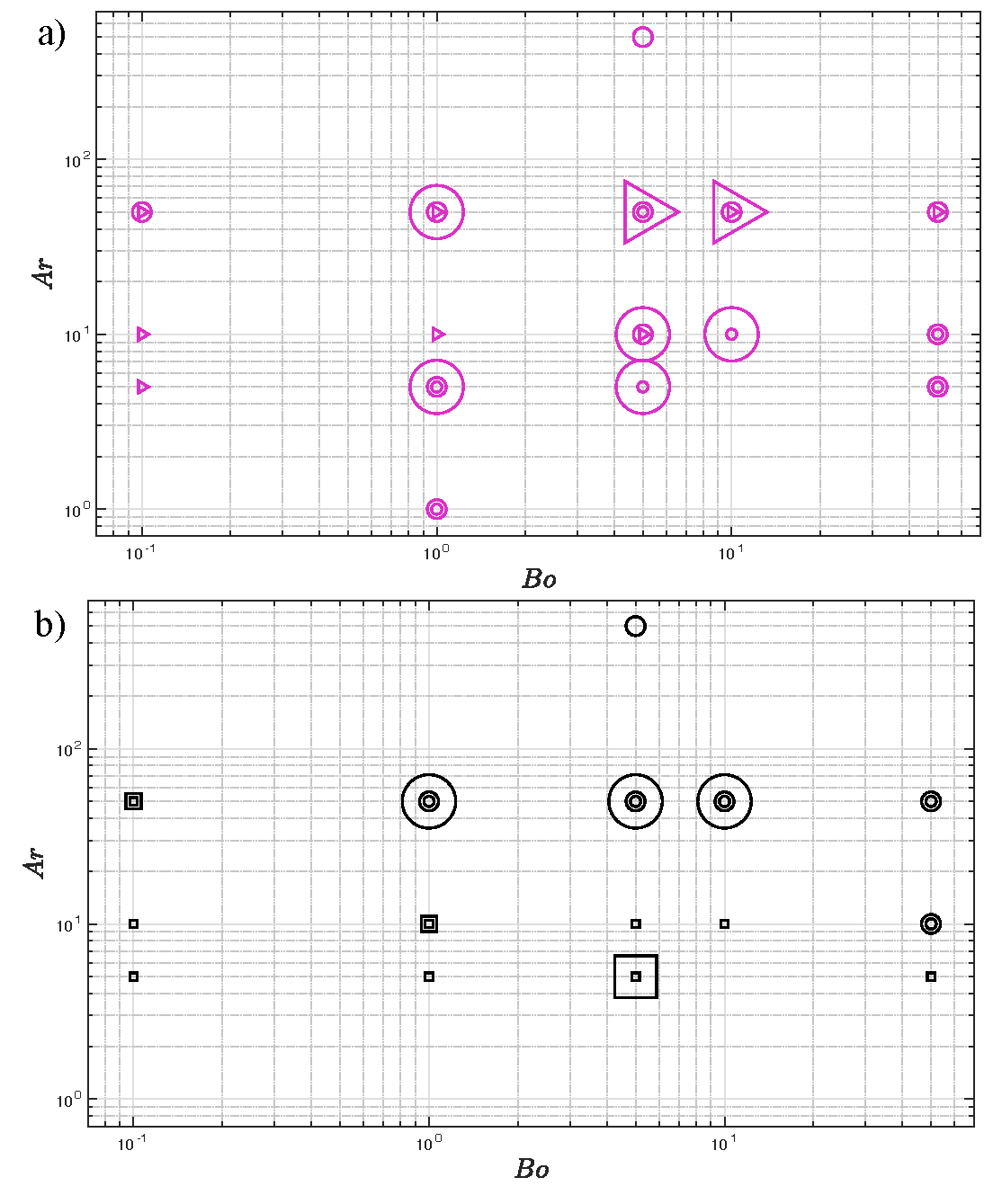}
\caption{Classification of entrainment regimes over the Ar-Bo domain for (a) $ \rho=0.7 $ and $ Y=0 $. (b) $ \rho=0.7 $ and $ Y=0.1 $. The size of symbols represents the value of $ m $ at each point, i.e $ m=0.1,~1,~10 $. Regimes are represnted by symbols as follows: $\square$ \textit{Plateau}, $\circ$ \textit{Descending - steady state}, $\triangleright$ \textit{Descending}, and $\diamond$ \textit{Ascending}.}
\label{fig:Regimesrho0_7}
\end{figure}

The entrainment regimes for $ \rho=0.7, Y=0 $ are shown in Fig.~\ref{fig:Regimesrho0_7}(a). In this case \textit{Descending} and \textit{Descending - steady state} regimes occur. Due to buoyancy, the entrained liquid is driven to settle downward. when $ m \leq 1 $, the settlement is faster. At low $ m $, the entrained liquid experiences less viscous shear stresses, also due to break up of the bubble upon entering the upper layer and formations of small bubbles, it takes longer to reach a steady state $ V_e $. We observe \textit{Descending} regime herein. At large $ m $, the entrained liquid experiences larger shear stresses in the upper layer. Combined with the negative bouyancy force $ V_e $ reduces but reaches a plateau, i.e. \textit{Descending - steady state}.

The entrainment regimes for $ \rho=0.7, Y=0.1 $ are shown in Fig.~\ref{fig:Regimesrho0_7}(b). In this case \textit{Plateau} and \textit{Descending - steady state} regimes occur. The entrained liquid descends and settles down due to negative buoyancy forces that it experiences. \textit{Descending - steady state} regime occurs when Bingham liquid has been trapped within the wake behind the rising bubble. The \textit{Plateau} regime occurs when there is no liquid trapped in the wake of the bubble.

The \textit{Plateau} regime is of particular interest since the Bingham liquid that has been pulled up into the lighter Newtonian liquid does not settle. This suggests that the buoyancy stresses are not strong enough to surpass the yield stress of the material, hence it remains in the upper layer.

\section{Maximum volume of entrainment $ V_e $}

These four regimes above primarily emphasize how far up from the liquid-liquid interface the bubbles can transport the lower layer fluid. In all regimes except for \textbf{ascending} regime, the maximum volume of entrained liquid has been transported cumulatively. However, there is no direct link between the entrainment regimes and the maximum volume of entrainment.

\begin{figure}
\hspace*{-0.1\textwidth}
\centering
\includegraphics[width=\textwidth, angle=0,clip=true, scale=1.2]{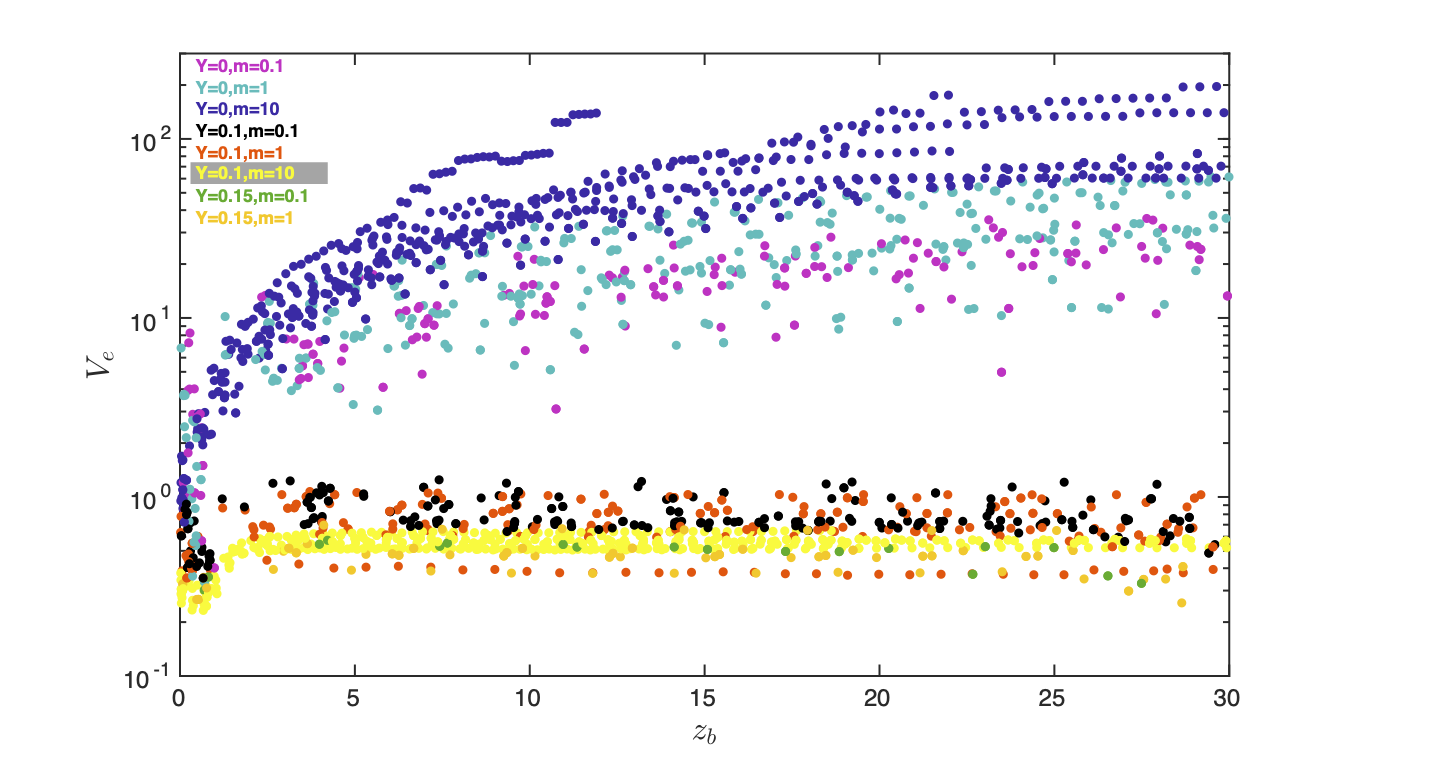}
\caption{ The volume of entrained fluid ($ V_e $) against the position of the bubble, where the fluid layers have similar densities ($ \rho=1 $). }
\label{fig:Ve}
\end{figure}

\begin{figure}
\hspace*{-0.1\textwidth}
\centering
\includegraphics[width=\textwidth, angle=0,clip=true, scale=1.2]{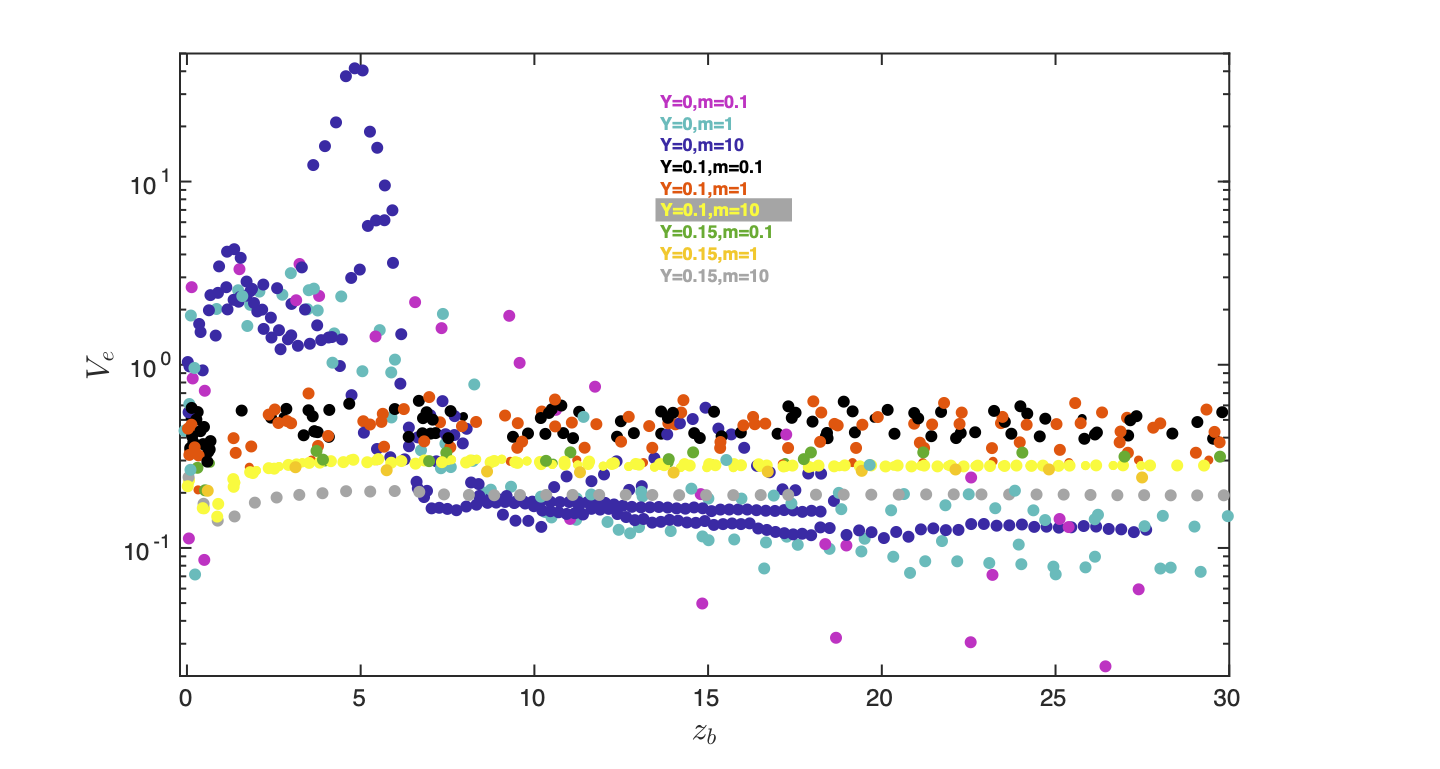}
\caption{ The volume of entrained fluid ($ V_e $) against the position of the bubble, where the upper fluid layer has less density than the lower fluid layer ($ \rho=0.7 $).}
\label{fig:VeHL}
\end{figure}

It is of interest to identify the set of flow parameters for which most liquid is transported. We have estimated the entrainment volume by integrating over the domain above $ z=0 $, including entrained fluid. This value has been normalized with the volume of the rising bubble. Variation of the volume of entrained fluid as it rises, for $ \rho=1 $, and $ \rho=0.7 $, is shown in Figs. \ref{fig:Ve} and \ref{fig:VeHL}, respectively. 
When the upper layer and the lower layer have the same density, ($ \rho=1 $), the maximum $ V_e $ occurs when both liquid layers are Newtonian. When the lower layer is a Bingham fluid, $ V_e $ decreases significantly, such that a bubble transports at most its own volume.

In the Newtonian case, the amount of liquid that has been transported decreases mainly by reducing  $ m $, i.e. the maximum entrainment observed occurs when $ m=10 $ and it reduces by reducing $ m $. By increasing $ m $ from 0.1 to 10, the viscosity of the fluid in the upper layer increases. The higher viscosity causes the column of entrained liquid to break off earlier. However, a larger area of the liquid-liquid interface has been displaced from the horizontal line ($ z=0 $). In addition, for $ m<1 $ and by increasing $ Ar $ and $ Bo $, a wake forms behind the rising bubble and some liquid becomes trapped there. The results show that the displaced fluid at the interface constitutes the more significant part of the entrained volume. At each $ m $, intermediate $ Bo $ values result more entrainment, and increasing $ Ar $ decreases the entrainment.

In the Bingham case, when $ m < 1 $ the bubble transports more fluid from the lower layer in comparison with $ m >1 $. The max value is quickly attained and generally remain constant (\textit{Plateau} regime). At each $ m $, decreasing $ Bo $ values and increasing $ Ar $ results in higher $ V_e $. For $ \rho=0.7$, similar to $ \rho=1 $, the maximum $ V_e $ happens for the Newtonian case with $ m=10 $. Although the density difference causes an appreciable reduction in the displacement of the interface, it is still comparable to other values of $ m $. Lastly, $ V_e $ does not diminish by having a Bingham fluid in the lower layer and a light Newtonian fluid in the upper layer.

\section{Summary}\label{sec:summary}

Dynamics of single bubbles rising from a viscoplastic fluid layer to a Newtonian fluid layer, and displacment of fluid from the lower layer to the upper layer has been studied computationally. The difference between the density and (effective) viscosity of the fluid layers is investigated and the results were compared with the Newtonian counterpart. The bubble is initially positioned in the viscoplastic fluid below a Newtonian fluid. As the bubble rises up it can undergoe breakup within the yield stress layer, with a small portion of it remaining confined within this layer. This phenomenon is found to occur for $ Ca \geqslant 5 $. Our results show that bubbles take on a prolate shape when passing through the viscoplastic-Newtonian interface at $ Bo >1 $, and it is more pronounced at $ m\leqslant 1 $. 

%Certain small-scale occurrences, like the later phases of pinch-off event, might not be adequately reproduced in computational models due to constraints in spatial resolution. However, this might not be a critical concern, as these small-scale features do not directly influence the subsequent stages of the overall dynamics.

Bubbles transport material from the lower layer to the upper layer through various mechanisms: displacing the liquid-liquid interface, forming a towed column of the lower layer liquid,and/or entrapping it within its wake. In cases where the lower layer is a Bingham fluid, the yield stress has been found to be sufficiently high to prevent the horizontal interface from being significantly displaced from its initial position as the bubble crosses it. This reduces the entrained volume. In general, the maximum $ V_e $ is found when the lower layer is Newtonian.

As the bubble ascends through the upper layer, the entrainment volume undergoes changes over time, resulting in four distinct entrainment regimes: \textit{Plateau}, \textit{Descending - steady state}, \textit{Descending}, and \textit{Ascending} regimes. On a larger scale, these regimes may imply that rising bubbles could create mixing regions (of fluids from two layers) at different depths. In the \textit{Plateau} regime, the mixing region is situated directly above the liquid-liquid interface and has a thickness of around 5 radii. A small fraction of the entrained liquid is likely to remain attached to the bubble and travels to the free surface. In the \textit{Ascending} regimes, bubbles have the capacity to transport the entrained material over greater distances and form thicker mixing layers above the liquid-liquid interface. In the case of the Bingham lower layer, the practical implication on a large scale is that the entrained fluid could either remain above the interface, forming localized mixing regions in the \textit{Plateau} regime, or be transported thigher and even reach the free surface.

It is found that the density difference of $ 70\% $ is not enough to prevent the entrainment of a yield stress fluid, $ Y=0.1 $. Although the entrainment decreases in comparison with $ \rho=1 $, there is still a possibility for formation of localized mixing region of upper and lower fluids due to entrainment. Moreover, the entrained heavy yield stress fluid doesn't recede over time, i.e. $ V_e $ for the case of $ \rho=0.7 $ and $ Y=0.1 $ does not change after the bubble tail pinches off. It is because the denser yield stress fluid can support itself while static.

\bibliographystyle{jfm}
\bibliography{biblio}

\end{document}